\documentclass[fleqn,usenatbib]{mnras}
\usepackage{multicol}
\usepackage{newtxtext,newtxmath}

\usepackage[T1]{fontenc}
\usepackage{float}
\usepackage[normalem]{ulem}
\DeclareRobustCommand{\VAN}[3]{#2}
\let\VANthebibliography\thebibliography
\def\thebibliography{\DeclareRobustCommand{\VAN}[3]{##3}\VANthebibliography}

\usepackage{graphicx}	
\usepackage{amsmath}	

\renewcommand{\vec}[1]{ {\bf #1} }

\title[High-order DG with sub-cell shock capturing]{High-order Discontinuous Galerkin hydrodynamics with sub-cell shock capturing on GPUs}

\author[M. Cernetic et al.]{%
Miha Cernetic$^{1}$\thanks{E-mail: cernetic@mpa-garching.mpg.de},
Volker Springel$^{1}$,
Thomas Guillet$^{2}$,
and R\"udiger Pakmor$^{1}$
\\%
$^{1}$Max Planck Institut f\"ur Astrophysik, Karl-Schwarzschild-Straße 1, 85748 Garching bei M\"unchen, Germany\\%
$^{2}$Physics and Astronomy, University of Exeter, Exeter EX4 4QL, UK
}

\date{Accepted 2023 April 04. Received 2023 April 01; in original form 2022 August 23}

\pubyear{2022}

\begin{document}
\label{firstpage}
\pagerange{\pageref{firstpage}--\pageref{lastpage}}
\maketitle

\begin{abstract}
Hydrodynamical numerical methods that converge with high-order hold particular promise for astrophysical studies, as they can in principle reach prescribed accuracy goals with higher computational efficiency than standard second- or third-order approaches. Here we consider the performance and accuracy benefits of Discontinuous Galerkin (DG) methods, which offer a particularly straightforward approach to reach extremely high order. Also, their computational stencil maps well to modern GPU devices, further raising the attractiveness of this approach. However, a traditional weakness of this method lies in the treatment of physical discontinuities such as shocks. We address this by invoking an artificial viscosity field to supply required dissipation where needed, and which can be augmented, if desired, with physical viscosity and thermal conductivity, yielding a high-order treatment of the Navier-Stokes equations for compressible fluids. We show that our approach results in sub-cell shock capturing ability, unlike traditional limiting schemes that tend to defeat the benefits of going to high order in DG in problems featuring many shocks. We demonstrate exponential convergence of our solver as a function of order when applied to smooth flows, such as the Kelvin-Helmholtz reference problem of Lecoanet et al. (2016). We also demonstrate excellent scalability of our GPU implementation up to hundreds of GPUs distributed on different compute nodes. In a first application to driven, sub-sonic turbulence, we highlight the accuracy advantages of high-order DG compared to traditional second-order accurate methods, and we stress the importance of physical viscosity for obtaining accurate velocity power spectra.    
\end{abstract}




\section{Introduction}

Computational fluid dynamics has become a central technique in modern astrophysical research \citep[for reviews, see, e.g.,][]{Trac2003, Vogelsberger2020, Andersson2021}. It is used in numerical simulations to advance the understanding of countless systems, ranging from planet formation \citep[e.g.][]{Nelson2000} over the evolution of stars \citep[e.g.][]{Edelmann2019}, and the interplay of gas, black holes and stars in galaxy formation \citep[e.g.][]{Weinberger2017}, up to extremely large scales involving clusters of galaxies \citep[e.g.][]{Dolag2009} or the filaments in the cosmic web \citep[e.g.][]{Mandelker2019}. 

This wide breadth of scientific applications is also mirrored in a bewildering diversity of numerical discretization schemes. Even so the underlying equations for thin, non-viscous gases -- the Euler equations -- are the same in a broad class of astrophysical studies, the commonly applied numerical methods come in many different flavors, and are sometimes based on radically different principles. At a basic level, one often distinguishes between Lagrangian and Eulerian discretization schemes. The former partition the gas into elements of (nearly) constant mass, as done for example in the popular smoothed particle hydrodynamics (SPH) approach \citep[e.g.][]{Monaghan1992} and its many derivatives. In contrast, the latter discretize the volume using a stationary (often Cartesian) mesh \citep[e.g.][]{Stone1992}, such that the fluid is represented as a field. Hybrid approaches, which for example use an unstructured moving-mesh \citep{Springel2010} are also possible. 

For mesh-based codes, finite-volume and finite-element methods are particularly popular. In the finite-volume approach, one records the averaged state in a cell, which is updated in time by the numerical scheme. This approach combines particularly nicely with the conservative character of the Euler equations, because the updates of the conserved quantities in each cell can be expressed as pair-wise fluxes through cell boundaries, yielding not only a manifestly conservative approach but also a physically intuitive formulation of the numerical method. In finite-element approaches one instead expands the fluid state in terms of basis functions. In spectral methods, the support of the basis functions can be the full simulation domain, for example if Fourier series are used to represent the system.

Discontinuous Galerkin (DG) approaches \citep[first introduced for non-linear problems by][]{Cockburn1989}, which are the topic of this paper, are a particular kind of finite-element approaches in which a series expansion for the solution is carried out separately within each computational cell (which can have a fairly general shape). Inside a cell, it is thus simply a truncated spectral method. The solutions for each of the cells are coupled with each other, however, at the surfaces of the cells. Interestingly, high-order accuracy of global solutions can be obtained simply through the high order of the spectral method applied inside a cell, while it does not require continuity of the solutions at the cell interfaces. This makes it particularly straightforward to extend DG schemes to essentially arbitrarily high order, because this does not make the coupling at cell interfaces any more complicated. This is quite different from high-order finite volume schemes, where the reconstruction step requires progressively deeper stencils at high order \citep{Janett2019}.

Another advantage of the DG approach is that it allows in principle cells of different convergence order to be directly next to each other \citep{Schaal2015}. This makes a spatially varying mesh resolution, or a spatially varying expansion order, more straightforward to implement than in high-order extensions of finite volume methods, where typically the high-order convergence property is compromised at resolution changes unless preserved with special treatments.

Despite these advantages, DG methods have only recently begun to be considered in astrophysics. First implementations and applications include \citet{Mocz2014, Schaal2015, Kidder2017, velasco_romero2018, guillet_high_order_2019}, as well as more recently \citet{Lombart2021, Markert2022, Deppe2022}. We here focus on exploring a new implementation of DG that we developed from the ground up for use with graphical processing units (GPUs). The recent advent of exascale supercomputers has been enabled through the use of graphical processing units (GPUs) or various other types of accelerator units. The common feature of these accelerators is the capability to execute a large number of floating point operations at the expense of lower memory bandwidth and total memory per computing unit (few MBs compared to few GBs on an ordinary compute node) compared to the CPU. Another peculiarity of accelerators is that they have hundreds of computing units (roughly equivalent to CPU cores) which execute operations in a single instruction, multiple data (SIMD) mode. Since many of the newest and largest supercomputers use such accelerators, it becomes imperative to either modify existing simulation codes for their efficient use, or to write new codes optimized for this hardware from scratch.

While there are already many successes in the literature for both approaches \citep[e.g.][]{Schneider2015, Ocvirk2016, Wibking2022},  most current simulation work in the astrophysical literature is still being carried out with CPU codes. Certainly one reason is that large existing code bases are not easily migrated to GPUs. Another is that not all numerical solvers easily map to GPUs, making it hard or potentially impossible to port certain simulation applications to GPUs.

However, there are also numerous central numerical problems where GPU computing should be applicable and yield sizable speed-ups. One is the study of hydrodynamics with uniform grid resolutions, as needed for turbulence. In this work, we thus focus on developing a new implementation of DG that is designed to run on GPUs. We base our implementation of DG on \citet{Schaal2015} and \citet{guillet_high_order_2019}, with one critical  difference. We do not apply the limiting schemes described in these studies as they defeat the benefits of high-order approaches when strong shocks are present. Rather, we will revert to the idea of deliberately introducing a small amount of artificial viscosity to capture shocks, i.e.~to add required numerical viscosity just where it is needed, and ideally with the smallest amount necessary to suppress unphysical oscillatory solutions. As we will show, with this approach the high-order approach can still be applied well to problems involving shocks, without having to sacrifice all high-order information on the stake of a slope limiter. 

This paper is structured as follows. In Section~\ref{SecEuler}, we detail the mathematical basis of the Discontinuois Galerkin discretization of hydrodynamics as used by us. In Section~\ref{SecViscosity}, we generalize the treatment to include source terms which involve derivatives of the fluid states, such as needed for the Navier-Stokes equations, or for our artificial viscosity treatment for that matter. We then turn to a discussion of shock capturing and oscillation control in Section~\ref{SecImplementation}. The following Section~\ref{SecBasicTests} is devoted to elementary tests, such as shock tubes and convergence tests for smooth problems. In Section~\ref{SecKH} we then show results for ``resolved'' Kelvin-Helmholtz instabilities, and in Section~\ref{SecTurbulence}, we give results for driven isothermal turbulence and discuss to what extent DG methods improve the numerical accuracy and efficiency of such simulations. Implementation and parallelization issues of our code, in particular with respect to using GPUs, are described in Section~\ref{SecCode}, while in Section~\ref{SecPerformance}, we discuss the performance and scalability of our new GPU-based hydrodynamical code. Finally, we give a summary and our conclusions in Section~\ref{SecSummary}.

\section{Discontinuous Galerkin discretization of the Euler equations}
\label{SecEuler}

The Euler equations are a system of hyperbolic partial differential equations. They encapsulate the conservation laws for mass, momentum and total energy of a fluid, and can be expressed as
\begin{equation}
    \label{eq:euler_equation}
    \frac{\partial \boldsymbol{u}}{\partial t}+\sum_{\alpha=1}^{d} \frac{\partial \boldsymbol{f}_{\alpha}(\boldsymbol{u})}{\partial x_{\alpha}}=0 ,
\end{equation}
where the sum runs over the $d$ dimensions of the considered problem. The state vector $\textbf{u}$ holds the conserved variables: density, momentum density, and total energy density:
\begin{equation}
    \label{eq:euler}
    \boldsymbol{u}=\left[\begin{array}{c}
\rho \\
\rho \boldsymbol{v} \\
e \\
\end{array}\right], \quad e=\rho u+\frac{1}{2} \rho \boldsymbol{v}^{2}.
\end{equation}

To make our system complete we need an equation of state which connects the hydrodynamics pressure $p$ with the specific internal energy $u$. If $\gamma$ is the adiabatic index, i.e.~the ratio of the specific heat of the gas at a constant pressure $C_p$ to its specific heat at a constant volume $C_v$, the ideal gas equation of state is
\begin{equation}
    p = \rho u \left( \gamma - 1 \right).
\end{equation}

We also need to specify the second term of Eq.~(\ref{eq:euler_equation}). The fluxes $\boldsymbol{f_\alpha} (\boldsymbol{u} )$ in three dimensions are:
\begin{equation}
    \label{eq:flux_matrix}
    \boldsymbol{f}_{1}=\left(\hspace*{-0.2cm}
    \begin{array}{c}
        \rho v_{x} \\
        \rho v_{x}v_{x}+p \\
        \rho v_{x} v_{y} \\
        \rho v_{x} v_{z} \\
        (\rho e+p) v_{x}
    \end{array}\hspace*{-0.2cm}\right), \; \boldsymbol{f}_{2}=\left(\hspace*{-0.2cm}
    \begin{array}{c}
        \rho v_{y} \\
        \rho v_{x} v_{y} \\
        \rho v_{y}v_{y} +p \\
        \rho v_{y} v_{z} \\
        (\rho e+p) v_{y}
    \end{array}\hspace*{-0.2cm}\right), \; \boldsymbol{f}_{3}=\left(\hspace*{-0.2cm}
    \begin{array}{c}
    \rho v_{z} \\
    \rho v_{x} v_{z} \\
    \rho v_{y} v_{z} \\
    \rho v_{z}v_{z}+p \\
    (\rho e+p) v_{z}
    \end{array}\hspace*{-0.2cm}\right).
\end{equation}
By summarizing the flux vectors into $\boldsymbol{F} = (\boldsymbol{f}_1, \boldsymbol{f}_2, \boldsymbol{f}_3)$, we can also write the Euler equations in the compact form
\begin{equation}
\label{eq:euler_compact}
\frac{\partial \boldsymbol{u}}{\partial t} +
\boldsymbol{\nabla} \cdot \boldsymbol{F} = 0, 
\end{equation}
which highlights their conservative character. Numerically solving this set of non-linear, hyperbolic partial differential equations is at the heart of computational fluid dynamics. Here we shall consider the specific choice of a high-order Discontinuous Galerkin (DG) method.

\subsection{Representation of conserved variables in DG}

In the Discontinuous Galerkin approach, the state vector $\boldsymbol{u}^K(\boldsymbol{x}, t )$ in each cell $K$ is  expressed as a linear combination of
time-independent, differentiable basis functions $\phi_l^K(\boldsymbol{x})$,
\begin{equation}
\label{eq:state_vector_expansion}
\boldsymbol{u}^{K}(\boldsymbol{x}, t) = \sum_{l=1}^{N} \boldsymbol{w}_{l}^{K}(t)\, \phi_{l}^{K}(\boldsymbol{x}),
\end{equation}
where the $\boldsymbol{w}_l^K(t)$ are  $N$ time dependent weights. Since the expansion is carried out for each component of our state vector separately, the weights $\boldsymbol{w}_l^K$ are really vector-valued quantities with 5 different values in 3D for each basis $l$. Each of these components is a single  scalar function with support in the cell $K$.

The union of cells forms a non-overlapping tessellation of the simulated domain, and the global numerical solution is fully specified by the set of all weights. Importantly, no requirement is made that the piece-wise smooth solutions within cells are continuous across cell boundaries. 

We shall use a set of orthonormal basis functions that is equal in all cells (apart from a translation to the cell's location), and we specialize our treatment in this paper to Cartesian cells of constant size. The DG approach can however be readily generalized to other mesh geometries, and to meshes with variable cell sizes. Also, we will here use a constant number $N$ of basis functions that is equal for all cells, and determined only by the global order $p$ of the employed scheme. In principle, however, DG schemes allow this be varied from cell to cell (so-called $p$-refinement).

\subsection{Time evolution}
\label{sec:time_stepping}

To derive the equations governing the time evolution of the DG weights $w_l^K$, we start with the original Euler equation from Eq.~(\ref{eq:euler_compact}), multiply it with one of the basis functions and integrate over the corresponding  cell $K$:
\begin{equation}
\int_K  \phi_{l}^{K}    \frac{\partial \boldsymbol{u}}{\partial t} {\rm d}\boldsymbol{x} +
\int_K
\phi_{l}^{K} \,\boldsymbol{\nabla} \boldsymbol{F} \, {\rm d}\boldsymbol{x} = 0. 
\end{equation}
Integration by parts of the second term and applying the divergence theorem leads to the so-called weak formulation of the conservation law:
\begin{equation}
\label{eq:weakform}
\int_K  \phi_{l}^{K}    \frac{\partial \boldsymbol{u}}{\partial t} {\rm d}\boldsymbol{x} +
\int_{\partial K}
\phi_{l}^{K} \, \boldsymbol{F} \, {\rm d}\boldsymbol{n}  
-
\int_K
\boldsymbol{\nabla} \phi_{l}^{K} \, \boldsymbol{F} \, {\rm d}\boldsymbol{x} = 0. 
\end{equation}
where $|K|$ stands for the volume of the cell (or area in 2D).

If we now insert the basis function expansion of $\boldsymbol{u}$ and make use of the orthonormal property of our set of basis functions,
\begin{equation}
    \int_K \phi_{l}^{K}(\boldsymbol{x}) \phi_{m}^{K}(\boldsymbol{x}) {\rm d}\boldsymbol{x} = \delta_{l, k} |K|,
\end{equation}
we obtain a differential equation for the time evolution of the weights:
\begin{equation}
\label{eq:weight_evolution}
|K| \frac{{\rm d} \boldsymbol{w}^K_{l}}{{\rm d} t}
= \int_K
\boldsymbol{\nabla} \phi_{l}^{K} \, \boldsymbol{F} \, {\rm d}\boldsymbol{x}
- \int_{\partial K}
\phi_{l}^{K} \, \boldsymbol{F}^\star(\boldsymbol{u}^+, \boldsymbol{u}^-) \, {\rm d}\boldsymbol{n}.
\end{equation}
Here we also considered that the flux function at the surface of cells is not uniquely defined if the states that meet at cell interfaces are discontinuous. We address this by replacing $\boldsymbol{F}(\boldsymbol{u})$ on cell surfaces with a flux function 
    $\boldsymbol{F}^{\star}(\boldsymbol{u}^+, \boldsymbol{u}^-)$ 
that depends on both states at the interface, where $\boldsymbol{u}^+$  is the outwards facing state relative to $\boldsymbol{n}$ (from the neighbouring cell),  and $\boldsymbol{u}^-$ is the state just inside the cell.  We will typically use a Riemann solver for determining $\boldsymbol{F}^{\star}$, making this akin to  Godunov's approach in finite volume methods. In fact, the same type of exact or approximate Riemann solvers can be used here as well. We use for ordinary gas dynamics a simplified version of the Riemann HLLC solver by \citet{toroRiemannSolvers} as implemented in the {\small AREPO} code \citep{Springel2010, Weinberger2020}. We have also included an exact Riemann solver in case an isothermal equation of state is specified.

What remains to be done to make an evaluation of Eq.~(\ref{eq:weight_evolution}) practical is to approximate both the volume and surface integrals numerically, and to choose a specific realization for the basis functions. We shall briefly discuss both aspects below. Another ingredient is the definition of the weights for the initial conditions. Thanks to the completeness of the basis, they can be computed by projecting the state vector $\boldsymbol{u}(\boldsymbol{x})$ of the initial conditions onto the basis functions $\phi_l^{K}$ of each cell:
\begin{equation}
\label{eq:weights_equation}
\boldsymbol{w}_{l}^{K} =  \frac{1}{|K|} \int_{K} \boldsymbol{u}\, \phi_{l}^{K} \mathrm{~d} V.
\end{equation}
If a finite number $N$ of basis functions is used to approximate the numerical solution, the total approximation error is then
\begin{equation}
\label{eq:l1_norm}
L1 =  \frac{1}{|K|} \int_{K} \left|\, \boldsymbol{u}(\boldsymbol{x})  -   \sum_{l=1}^{N} \boldsymbol{w}_{l}^{K}\, \phi_{l}^{K}(\boldsymbol{x})\, \right| \mathrm{~d} V.
\end{equation}
We shall use this L1 norm to examine the accuracy of our code when analytic solutions are known.

\begin{figure}
	\includegraphics[width=78mm]{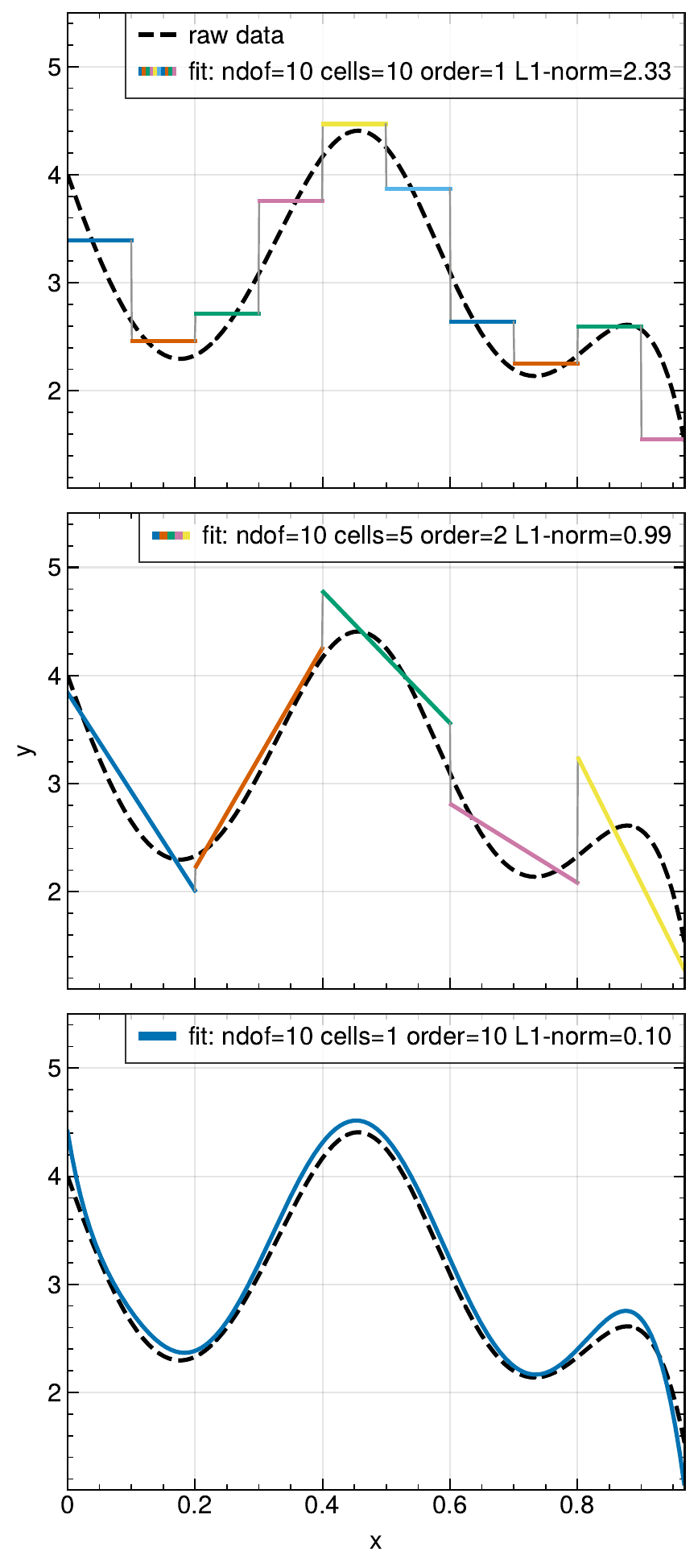}
    \caption{An example of fitting an arbitrary, smooth function $y=f(x)$ with 10 degrees of freedom, but varying number of cells and polynomial orders used for these cells, as labelled in the different panels. The L1 error norm for approximating the function is highest in case piece-wise constant approximations are used, while it drops when fewer, but piece-wise linear cells are used, and finally reaches its lowest value when a single cell with a single 10th order polynomial is used.}
    \label{fig:ndof_fitting_example}
\end{figure}

\subsection{Legendre basis function}

Following \citet{Schaal2015}, we select Legendre polynomials $P_l(\xi)$ to construct our set of basis functions. They are defined on a canonical interval $[-1,1]$ and can be scaled such that they form an orthogonal basis with normalization chosen as:
\begin{equation}
    \int_{-1}^{1} P_l(\xi) P_m(\xi) {\rm d}\xi = 2\, \delta_{l,m} .
\end{equation}
Note that the 0-th order Legendre polynomial is just a constant term, while the 1-st order features a simple pure linear dependence. In general, $P_l(\xi)$ is a polynomial of degree $l$.

Within each cell, we define local coordinates $\boldsymbol{\xi} \in \left[ -1, 1 \right]^d$. The translation between global coordinates $\boldsymbol{x}$ to local cell coordinates $\boldsymbol{\xi}$ is:
\begin{equation}
\label{eq:coordinate_translation}
\boldsymbol{\xi}^K=\frac{2}{h}\left(\boldsymbol{x}-\boldsymbol{x}_{c}^{K}\right),
\end{equation}
with $h$ being the cell size in one dimension, and $\boldsymbol{x}_c^K$ is the cell centre in world coordinates. Multi-dimensional basis functions are simply defined as Cartesian products of Legendre polynomials, for example in three dimensions as follows:
\begin{equation}
\label{eq:coord_trans}
\phi_{l}^{K} (\boldsymbol{x}) = P_l^{\rm 3D}[ \boldsymbol{\xi}^K(\boldsymbol{x}) ], 
\end{equation}
with
\begin{equation}
P_l^{\rm 3D}[ \boldsymbol{\xi}^K ] \equiv  P_{l_x}(\xi^K_x)\cdot P_{l_y}(\xi^K_y) \cdot P_{l_z}(\xi^K_z),
\end{equation}
where the generalized index $l$ enumerates different combinations of Legendre polynomials $l_x(l)$, $l_y(l)$, and $l_z(l)$ in the different directions. In practice, we truncate the expansion at a predefined order $n$, and discard all tensor products in which the degree of the resulting polynomial exceeds $n$.  This means that we end up in 3D with 
\begin{equation}
N^{\rm 3D}(n)=\frac{1}{6}(n+1)(n+2)(n+3)
\label{eqn:N3D}
\end{equation}
basis functions, each a product of three Legendre polynomials of orders   $l_{z,y,z}\in \{0, \ldots, n\}$. In 2D, we have
\begin{equation}
N^{\rm 2D}(n)= \frac{1}{2}(n+1)(n+2),
\label{eqn:N2D}
\end{equation}
and in 1D the number is $N^{\rm 1D}(n)= n +1$. The expected spatial convergence order due to the leading truncation error is in each case $p=n+1$. From now on we will refer to $p$ as the order of our DG scheme, with $n=p-1$ being the highest degree among the involved Legendre polynomials.

In Figure~\ref{fig:ndof_fitting_example}, we show an example of approximating a smooth function with Legendre polynomials of different order and with a different number of cells, but keeping the number of degrees constant. In this case, the approximation error tends to be reduced by going to higher order, even when this implies using fewer cells.

\subsection{Gaussian quadrature}

An integration of a general function $f(x)$ over the interval $[-1,1]$ can be approximated by Gaussian quadrature rules, as
\begin{equation}
\int_{-1}^{1} f(x)\, {\rm d} x \simeq \sum_{j=1}^{n_g} g_j\, f(x_j)
\end{equation}
for a set of evaluation points $x_j$ and suitably chosen quadrature weights $g_j$. We use ordinary Gaussian quadrature with internal points only. The corresponding integration rule with $n_g$ evaluation points is exact for polynomials up to degree $2 n_g -1$. If we use Legendre polynomials up to order $n$, we therefore should use at least $n_g \ge (n + 1)/2$ integration points. Note, however, that the nonlinear dependence of the flux function on the state vector $\boldsymbol{u}$ means that we actually encounter rational functions as integrands and not just simple polynomials. As a result, we need unfortunately a more conservative number of integration points for sufficient accuracy and stability in practice. A good heuristic is to take the number of basis functions used for the one-dimensional case as a guide, so that one effectively employs at least one function evaluation per basis function.  This means we pick  $n_g = n+1$ in what follows.

Multi-dimensional integrations, as needed for the  surface and volume integrals in our Cartesian setup, can be carried out through tensor products of Gaussian integrations. We denote the corresponding function evaluation points as $\boldsymbol{\xi}^{\rm vol}_{\boldsymbol{j}} = (x_{j_1}, x_{j_2}, x_{j_3})$ and Gaussian weights as $g^{\rm vol}_{\boldsymbol{j}} = g_{j_1} \cdot g_{j_2} \cdot g_{j_3}$ for the combination  $\boldsymbol{j} = (j_1, j_2, j_3)$ of Gaussian quadrature points needed for integrations over the cell volume in 3D. For  surface integrations over our cubical cells, we correspondingly define
$\boldsymbol{\xi}^{\rm sur}_{\boldsymbol{k}, x+} = (+1, x_{k_1}, x_{k_2})$, and $\boldsymbol{\xi}^{\rm sur}_{\boldsymbol{k}, x-} = (-1, x_{k_1}, x_{k_2})$ for evaluation points on the right and left surface in the $x$-direction of one of our cubical cells, with $\boldsymbol{k} = (k_1, k_2)$ and likewise for the $y$- and $z$-directions. The corresponding Gaussian quadrature weights are given by $g^{\rm sur}_{\boldsymbol{k}} = g_{k_1} \cdot g_{k_2}$.

Putting everything together, we arrive at a full set of discretized evolutionary equations for the weights.
For definiteness, we specify this here for the three dimensional case:
\begin{align}
\label{eq:time_stepping_final_analytical_equation}
\frac{{\rm d} \boldsymbol{w}^K_{l}}{{\rm d} t} =
\frac{1}{4}
\sum_{\alpha=1}^{3}
\sum_{\substack{\boldsymbol{j}   \in  \\ [1,n_g]^3}}
\left\{\boldsymbol{f}_{\alpha}[\boldsymbol{u}^{K}(\boldsymbol{\xi}^{\rm vol}_{\boldsymbol{j}})] \cdot \frac{\partial P_l^{\rm 3D}(\boldsymbol{\xi}^{\rm vol}_{\boldsymbol{j}})}{\partial \xi_{\alpha}}\right\}
g^{\rm vol}_{\boldsymbol{{j}}}  \nonumber \\
-\frac{1}{8}
\sum_{\alpha=1}^{3}
\sum_{\substack{\boldsymbol{k}\in \\ [1,n_g]^2}}
\Bigg\{ P_l^{\rm 3D}(\boldsymbol{\xi}^{\rm sur}_{\boldsymbol{k},\alpha+})\,   \boldsymbol{f}^{\star}_\alpha\left[\boldsymbol{u}^{K,\alpha+}(\xi^{\rm sur}_{\boldsymbol{k},\alpha -}),  \boldsymbol{u}^{K}(\boldsymbol{\xi}^{\rm sur}_{\boldsymbol{k},\alpha +})\right]   
  \nonumber \\
- P_l^{\rm 3D}(\boldsymbol{\xi}^{\rm sur}_{\boldsymbol{k},\alpha+})\,   \boldsymbol{f}^{\star}_\alpha\left[  \boldsymbol{u}^{K}(\boldsymbol{\xi}^{\rm sur}_{\boldsymbol{k},\alpha -}),
\boldsymbol{u}^{K,\alpha-}(\boldsymbol{\xi}^{\rm sur}_{\boldsymbol{k},\alpha+})\right]
\Bigg\}  g^{\rm sur}_{\boldsymbol{{k}}} .
\end{align}
Here the notation $\boldsymbol{u}^{K,\alpha+}$ and $\boldsymbol{u}^{K,\alpha-}$ refer to the state vectors evaluated for the right and left neighbouring cells of cell $K$ in the direction of axis $\alpha$, respectively. The state vector evaluations themselves are  given by 
\begin{equation}
    \boldsymbol{u}^{K}(\boldsymbol{\xi}) = \sum_{l=1}^{N} \boldsymbol{w}_{l}^{K}\, P_l^{\rm 3D}(\boldsymbol{\xi}).
\end{equation}
Note that the prefactor $1/|K|$ in front of the surface integral terms in Eq.~(\ref{eq:time_stepping_final_analytical_equation}) turns into $1/8$ as a result of the change of integration variables mediated by Eq.~(\ref{eq:coord_trans}). The volume integral acquires a factor of $2/h$ from the coordinate transformation, thus the final prefactor becomes $1/4$. The numerical computation of the time derivative of the weights based on a current set of weights is in principle straightforward using Eq.~(\ref{eq:time_stepping_final_analytical_equation}), but evidently becomes more elaborate at high-order, involving numerous sums per cell. 

In passing we note that instead of just counting the number of cells per dimensions, both the storage effort and the numerical work needed is better measured in terms of the number of degrees of freedom per dimension. A fixed number of degrees of freedom (and thus storage space) can be achieved with different combinations of cell size and expansion order. The hope in  using high-order methods is that they deliver better accuracy for a fixed number of degrees of freedom, or arguably even more importantly, better accuracy at fixed computational expense.  

\subsection{Time integration}
With
\begin{equation}
\boldsymbol{\dot{w}} \equiv \frac{{\rm d} \boldsymbol{w}^K_{l}}{{\rm d} t}
\end{equation}
 in hand, standard ODE integration methods such as the broad class of Runge-Kutta integrations can be used to advance the solution forward in time. We follow standard procedure and employ strongly positivity preserving (SPP) Runge-Kutta integration rules as defined in \citet[Appendix D]{Schaal2015}. Note that when higher spatial order is used, we correspondingly use a higher order time integration method, such that the time integration errors do not start dominating over spatial discretization errors. The highest time integration method we use is a 5 stage 4-th order SSP RK method.

The timestep size $\Delta t$ is set conservatively as
\begin{equation}
    \Delta t_\textrm{max} = f_\textrm{CFL} \frac{h}  {2p \left(c_{s\textrm{, max}} + v_\textrm{max} \right)},
    \label{eq:timestep}
\end{equation}
where $h$ is the cell size, $f_\textrm{CFL}$ is the Courant–Friedrichs–Lewy factor, $c_{s\textrm{, max}}$ denotes the global maximum sound speed and $v_\textrm{max}$ is the global maximum kinematic velocity, respectively. We use a $f_\textrm{CFL}$ of 0.5 for all problems except the shock tube, Sedov blast wave and double blast wave where a more conservative 0.3 was used instead.

For high order runs ($p>4$) we did not see time integration errors to start dominating over the spatial discretization errors, despite employing only a 4-th order RK scheme. We attribute this to our use of a low Courant factor and to including global maximum velocities in the timestep criterion. Once the errors from time integration would start to dominate at high order, we could recover sufficient accuracy of our time integration scheme by appropriately scaling the time-step size as $h^{r/4}$.

\section{Treatment of viscous source terms}
\label{SecViscosity}

As we will discuss later on, our approach for capturing physical discontinuities (i.e.~shocks and contact discontinuities) in gas flows deviates from the classical slope-limiting approach and instead relies on a localized enabling of artificial viscosity. Furthermore, we will generalize our method to also account for physical dissipative terms, so that we arrive at a treatment of the full compressive Navier-Stokes equations. 

To introduce these methods, we start with a generalized set of Euler equations in 3D that are augmented with a diffusion term in all fluid variables,
\begin{equation}
    \label{eq:navier-stokes-bassi}
    \frac{\partial \boldsymbol{u}}{\partial t}+\nabla \cdot \boldsymbol{F}=\nabla \cdot(\varepsilon \nabla \boldsymbol{u}),
\end{equation}
where $\boldsymbol{u}$ and $\boldsymbol{F}$ are the state vector (\ref{eq:state_vector_expansion}) and the flux matrix (\ref{eq:flux_matrix}), respectively.

The crucial difference between the normal Euler equations~(\ref{eq:euler_equation}) and this dissipative form is the introduction of  a second derivative on the right-hand side, which modifies the character of the problem from being purely hyperbolic to an elliptic type, while retaining manifest conversation of mass, momentum and energy. This second derivative can however not be readily accommodated in our weight update equation obtained thus far. Recall, the reason we applied integration by parts and the Gauss' theorem going from Eq.~(\ref{eq:euler_compact}) to Eq.~(\ref{eq:weakform}) was to eliminate the spatial derivative of the fluxes. If we apply the same approach to $\nabla \cdot(\varepsilon \nabla \boldsymbol{u})$ we are still left with one $\nabla$-operator acting on the fluid state. 

\subsection{The uplifting approach}

In a seminal paper, \citet{bassi_rebay_1997JCoPh.131..267B} suggested a particular treatment of this second derivative inspired by how one typically reduces second (or higher) order ordinary differential equations (ODEs) to first order ODEs.  \citet{bassi_rebay_1997JCoPh.131..267B} reduce the order of Eq.~(\ref{eq:navier-stokes-bassi}) by introducing the gradient of the state vector, $\boldsymbol{S} \equiv \nabla \boldsymbol{u}$, as an auxiliary set of unknowns. This yields a system of two partial differential equations:
\begin{align}
    \label{eq:uiss}
    &\boldsymbol{S} - \nabla \boldsymbol{u} = 0, \\
    \label{eq:navier-stokes-reduced-order}
    &\frac{\partial \boldsymbol{u}}{\partial t}+\nabla \cdot (\boldsymbol{F} - \epsilon\boldsymbol{S})=0.
\end{align}
Interestingly, if we consider a basis function expansion for $\boldsymbol{S}$ for each cell in the same way as done for the state vector, then the weak formulation of the first equation can be solved with the DG formalism using as input only the series expansion of the current state $\boldsymbol{u}$. This entails again an integration by parts that yields volume and surface integrations for each cell. To compute the latter, one needs to adopt a surface state $\boldsymbol{u}^\star$ for potentially discontinuous jumps $\boldsymbol{u}^+$ and $\boldsymbol{u}^-$ across the cell boundaries. \citet{bassi_rebay_1997JCoPh.131..267B} suggest to use the arithmetic mean $\boldsymbol{u}^\star = [\boldsymbol{u}^- + \boldsymbol{u}^+]/2$ for this, so that obtaining the series expansion coefficients for $\boldsymbol{S}$ is straightforward. One can then proceed to solve Eq.~(\ref{eq:navier-stokes-reduced-order}), with a largely identical procedure than for the Euler equation, except that the ordinary flux $\boldsymbol{F}$ 
is modified by subtracting the viscous flux $\boldsymbol{F}_{\rm visc} = \epsilon\boldsymbol{S}$. At cell interfaces one furthermore needs to define the viscous flux uniquely somehow, because $\boldsymbol{S}$ can still be discontinuous in general at cell interfaces. Here \citet{bassi_rebay_1997JCoPh.131..267B} suggest to use the arithmetic mean again.

A clear disadvantage of this procedure, which we initially implemented in our code, is that it significantly increases the computational cost, memory requirements and code complexity, because the computation of $\boldsymbol{S}$ involves the same set of volume and surface integrals that are characteristic of the DG approach, except that it actually has to be done {\em three times} as often than for $\boldsymbol{u}$ in 3D, once for each spatial dimension. But more importantly, we have found that this method is prone to robustness problems, in particular if the initial conditions already contain large discontinuities across cells. In this case, the estimated derivatives inside a cell can reach unphysically large values by the jumps seen on the outer sides of a cell. 

In hindsight, this is perhaps not too surprising. For a continuous solution, there is arguably little if anything to be gained by solving Eq.~(\ref{eq:uiss}) with the DG algorithm if a polynomial basis is in use. Because this must then return a solution identical to simply taking the derivatives of the basis functions (which are analytically known) and retaining the coefficients of the expansion. On the other hand, if there are discontinuities in $\boldsymbol{u}$ at the boundaries, the solution for $\boldsymbol{S}$ sensitively depends on the (to a certain degree arbitrary) choice made for resolving the jumps in the computation of the surface integrals for $\boldsymbol{S}$. In particular, there is no guarantee that using the arithmetic mean does not induce large oscillations or unphysical values for $\boldsymbol{S}$ in the interior of cells in certain cases.

For all these reasons we have ultimately abandoned the \citet{bassi_rebay_1997JCoPh.131..267B} method, because it does not yield a robust solution for the diffusion part or the equations in all situations, and does not converge rapidly at high order either. Instead, we conjecture that the key to high order convergence of the diffusive part of the PDE system is the availability of a consistently defined continuous solution across cell boundaries.

\subsection{Surface derivatives}

For internal evaluations of the viscous flux (which in general may depend on $\boldsymbol{u}$ and $\nabla\boldsymbol{u}$) within a cell, we use the current basis function expansion of the solution in the cell and simply obtain the derivative by analytically differentiating the basis functions. We argue that this is the most natural choice as the same interior solution $\boldsymbol{u}$ is used for computing the ordinary hydrodynamical flux. 

The problem, however, lies with the surface terms of the viscous flux, as here neither the value of the state vector nor the gradient are uniquely defined, and unlike for the hyperbolic part of the equation, there is no suitable `Riemann solver' to define a robust flux for the diffusion part of the equation. Simply taking arithmetic averages of the two values that meet at the interface for the purpose of evaluating the surface viscous flux is not accurate and  robust in practice.

\begin{figure}
\begin{center}
	\resizebox{5.0cm}{!}{\includegraphics{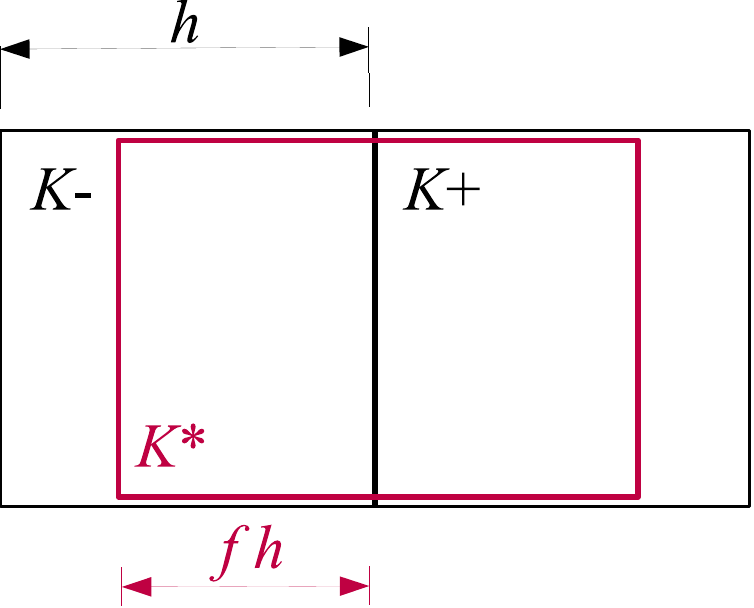}}
	\end{center}
    \caption{Two cells $K^-$ and $K^+$ that meet at a joint face. The corresponding polynomial solutions $u-$ and $u+$ are in general discontinuous at the interface. To unambiguously define a joint solution and its gradient on the interface, we construct an interpolant solution on a domain $K*$  placed symmetrically around the interface. In the normal direction, a fraction $f$ of both cells is covered (we pick either $f=3/4$ or $f=1$ in practice), in the the transverse direction(s), the cells are covered in full. 
    \label{fig:sketch_surface_projection}}
\end{figure}

We address this problem by constructing a new {\em continuous} solution across a cell interface by considering the current solutions in the two adjacent cells of the interface, and projecting them onto a new joint polynomial expansion in a rectangular domain that covers part (or all) of the two adjacent cells. This approach is similar to the recovery method proposed by \citet{dg_recovery_2005} in their work on solving the diffusion equation in DG. This interpolated solution minimizes the $L_2$ difference to the original (in general discontinuous) solutions in the two cells, but it is continuous and differentiable at the cell interface by construction. The quantities  $\boldsymbol{u}$ and $\nabla\boldsymbol{u}$ needed for the evaluation of the viscous surface flux are then computed by evaluating the new basis function expansion at the interface itself. 

A sketch of the adopted procedure is shown in Figure~\ref{fig:sketch_surface_projection}. The two solutions in the two adjacent cells are given by
  \begin{equation}
    \boldsymbol{u}^{K^-}(\boldsymbol{x}) = \sum_{l=1}^{N} \boldsymbol{w}_{l}^{K^-}\, \phi^{K^-}(\boldsymbol{x}).
\end{equation}
and
  \begin{equation}
    \boldsymbol{u}^{K^+}(\boldsymbol{x}) = \sum_{l=1}^{N} \boldsymbol{w}_{l}^{K^+}\, \phi^{K^+}(\boldsymbol{x}).
\end{equation}
We now seek an interpolated solution in terms of a set of new basis functions $\psi^{K^\star}$ defined on the domain $K^\star$, i.e.
  \begin{equation}
    \boldsymbol{\tilde{u}}^{K^\star}(\boldsymbol{x}) = \sum_{l=1}^{N^{\star}} \boldsymbol{q}_{l}^{K^\star}\, \psi^{K^\star}(\boldsymbol{x}).
\end{equation}
In order to avoid a degradation of accuracy if the solution is smooth, and to provide sufficient accuracy for the gradient, we adopt order $n+1$ for the polynomial basis of $\boldsymbol{\tilde{u}}^{K^\star}$. As for ordinary cells, the generalized index $l$ enumerates different combinations $[l_x(l), l_y(l), l_z(l)]$ of Legendre polynomials and their Cartesian products in the multidimensional case. If, for example, the two cells are oriented along the $x$-axis, we define
\begin{equation}
\label{eq:coord_trans2}
\psi_{l}^{K^\star} (\boldsymbol{x}) =  P_{l_x}(\xi_x^{K^\star})\cdot P_{l_y}(\xi^K_y) \cdot P_{l_z}(\xi^K_z),
\end{equation}
where now the mapping of the $x$-extension of the domain $K^\star$ into the standard interval $[-1,1]$ is correspondingly modified as
\begin{equation}
\label{eq:coordinate_translation_viscosity}
\xi_x^{K^\star}=\frac{1}{f\,h}\left({x}-\frac{x_{c}^{K^-} +  {x}_{c}^{K^+}}{2}\right),
\end{equation}
where $f$ is the fraction of overlap of each of the two cells (see Fig.~\ref{fig:sketch_surface_projection}). The coefficients $\boldsymbol{q}_{l}^{K^\star}$ can then be readily obtained by carrying out the projection integrals
\begin{eqnarray}
\boldsymbol{q}_{l}^{K^\star} & = & \frac{1}{|K^\star|}\int_{K^\star} \boldsymbol{u}(\boldsymbol{x})\; 
\psi_{l}^{K^\star}(\boldsymbol{x})\,{\rm d}V \\
&  \hspace*{-1.7cm}= & \hspace*{-1cm}\frac{1}{|K^\star|}\sum_{m=1}^{N} \left[ 
\boldsymbol{w}_m^{K^-} \int_{K^-} \phi_m^{K^-} \psi_{l}^{K^\star}  \, {\rm d}V  
+
\boldsymbol{w}_m^{K^+} \int_{K^+} \phi_m^{K^+} \psi_{l}^{K^\star} \, {\rm d}V  \nonumber
\right].
\end{eqnarray}
The projection is a linear operation, and the overlap integrals of the Legendre basis functions can be precomputed ahead of time. In fact,  many evaluate to zero due to the orthogonality of our Legendre basis. In particular, this is the case for the transverse basis functions if their order is not equal, so that the projection effectively becomes a sparse matrix operation that expresses the new expansion coefficients in the normal direction as a sum of one or several old expansion coefficients in the normal direction. This can be more explicitly seen by defining Legendre overlap integrals as
\begin{eqnarray}
    A_{m,l}^- &=& \int_{-1}^{0} P_m(2 f x + 1) P_l(x) \,{\rm d}x,\\
        A_{m,l}^+ &=& \int_{0}^{1} P_m(2 f x + 1) P_l(x) \,{\rm d}x.
\end{eqnarray}
Then the new coefficients can be computed as follows
\begin{equation}
\boldsymbol{q}_{(l_x,l_y,l_z)}^{K^\star}
= \frac{1}{2f}\sum_{m_x =0}^{l_x}
\left[ 
A_{m_x,l_x}^- \boldsymbol{w}_{(m_x, l_y, l_z)}^{K^-}  + A_{m_x,l_x}^+ \boldsymbol{w}_{(m_x, l_y, l_z)}^{K^+} 
\right].
\label{eqn:basisprojection}
\end{equation}
Note that for transverse dimensions, only the original Legendre polynomials contribute, hence the new coefficients are simply linear combinations of coefficients that differ only in the order of the Legendre polynomial in the $x$-direction. Also note that for the transverse dimensions, the highest Legendre orders $l_y$ and $l_z$ that are non-zero are the same as for the original coefficients, i.e.~the fact that we extend the order to $n+1$ becomes only relevant for the direction connecting the two cells.

Another point to note is that the basis function projection can be carried out independently for the left and right side of an interface (corresponding to the first and second part of the sum in eqn.~\ref{eqn:basisprojection}), each yielding a partial result that can be used in turn  to  evaluate partial results for $\boldsymbol{\tilde{u}}$ $\nabla\boldsymbol{\tilde{u}}$  at the interface. Adding  up these partial results then yields the final interface state and and interface gradient. This means that this scheme does not require to send the  coefficients $\boldsymbol{w}^{K^\pm}$ to other processors in case $K^-$ and $K^+$ happen to be stored on different CPUs or GPUs, only ``left'' and ``right'' states for $\boldsymbol{\tilde{u}}$ and $\nabla \boldsymbol{\tilde{u}}$ need to be exchanged (which are the partial results that are then summed instead of taking their average), implying the same communication costs as, for example, methods that would rely on taking arithmetic averages of the values obtained separately for the $K^-$ and $K^+$ sides.
  
Finally, we choose $f=3/4$ for the size of the overlap region for $n\le 2$, but $f=1$ for  higher order $n> 2$. For the choice of $f=3/4$, the estimate for the first derivative of the interpolated solution ends up being 
\begin{equation}
\nabla \boldsymbol{\tilde{u}} =  \frac{\boldsymbol{u}^{+} - \boldsymbol{u}^{-}}{h} \boldsymbol{n},
\end{equation}
for piece-wise constant states, 
where $h$ is the cell spacing, $\boldsymbol{n}$ is the normal vector of the interface, and $\boldsymbol{u}^\pm$ are the average states in the two cells. This intuitively makes sense for low order. In particular, this will pick up a reasonable gradient even if one starts with a piece-wise constant initial conditions, and even if $n=0$ (corresponding to DG order $p=1$) is used. We also obtain the expected convergence orders for diffusion problems (see below) with this choice when  $n \le 2$ is used. On the other hand, we have found that it is necessary to include the full available information of the two adjacent cells by adopting $f=1$ for still higher order in order to obtain the expected high-order convergence rates for diffusion problems also for $n > 2$.

\subsection{The Navier-Stokes equations}
\label{SecNavierStokesEquations}
While we will use the above form of the dissipative terms for our treatment of artificial viscosity (see below), we also consider the full Navier-Stokes equations. They are given by:
\begin{equation}
    \label{eq:navier-stokes-basic}
    \frac{\partial \boldsymbol{u}}{\partial t}+\nabla \cdot \boldsymbol{F}=\nabla \cdot  \boldsymbol{F}_{\rm NS},
\end{equation}
where now the Navier-Stokes flux vector $\boldsymbol{F}_{\rm NS}$ is a non-linear function both of the state vector $\boldsymbol{u}$  and its gradient $\boldsymbol{\nabla{u}}$. We pick the canonical form
\begin{equation}
 \boldsymbol{F}_{\rm NS} = \left( \begin{array}{c}
        0 \\
        \boldsymbol{\Pi} \\
        \boldsymbol{v}\cdot\boldsymbol{\Pi}  +  \chi (\gamma - 1) \rho \nabla u\\
    \end{array}
    \right) ,
\end{equation}
with a viscous tensor
\begin{equation}
    \boldsymbol{\Pi} = \nu \rho \left(
    \nabla{\boldsymbol{v}} + \nabla{\boldsymbol{v}}^{\rm T} + \frac{2}{3}\nabla\cdot \boldsymbol{v}\right)
\end{equation}
that dissipates shear motions with viscosity $\nu$. We also include optional heat conduction with thermal diffusivity $\chi$. Note that the derivatives of the primitive variables can be easily obtained from the derivatives of the conservative variables when needed, for example $\boldsymbol{\nabla v} = [\nabla(\rho \boldsymbol{v}) - \boldsymbol{v} \nabla\rho]/\rho$, and one can thus express the velocity gradient $\boldsymbol{\nabla v}$ in terms of  $\boldsymbol{\nabla{u}}$ and $\boldsymbol{u}$.

\subsection{Passive tracer}
\label{SecPassiveScalar}

Finally, for later application to the Kelvin-Helmholtz problem, we follow \citet{Lecoanet2016} and add a passive, conserved tracer variable to the fluid equations. The density of the tracer is $c\rho$, with $c$ being its dimensionless relative concentration. It can be added as a further row to the state vector $\boldsymbol{u}$. Since the tracer is conserved and simply advected with the local velocity, the corresponding entry in the flux vector is $c\rho\boldsymbol{v}$. Further, we can also allow for a diffusion of the tracer with diffusivity $\eta$, by adding $\eta \rho \nabla c$ in the corresponding row of the Navier-Stokes flux vector. The governing equation for the passive tracer dye is hence
\begin{equation}
\frac{\partial (c \rho)}{\partial t} +\nabla \cdot (c \rho\boldsymbol{v}) = \nabla(\eta \rho \nabla c).
\label{eqn:passivetracer}
\end{equation}

\section{Shock capturing and oscillation control}
\label{SecImplementation}

\subsection{Artificial viscosity}

High-order numerical methods are prone to oscillatory behaviour around sharp jumps of density or pressure. Such physical discontinuities arise naturally at shocks in supersonic fluid motion, and they are an ubiquitous phenomenon in astrophysical gas dynamics. In fact, the Euler equations have the interesting property that perfectly initial conditions can evolve with time into states that feature real discontinuities. The physical dissipation that must happen in these jumps is implicitly dictated by the conservation laws, but discrete numerical methods may not always produce the required level of dissipation, such that postshock oscillations are produced that are reminiscent of the Gibbs phenomenon in Fourier series expansion around jump discontinuities.

Our DG code produces these kinds of oscillations with increasing prominence at higher and higher order when discontinuities are present. And once the oscillations appear, they do not necessarily get quickly damped because of the very low numerical dissipation of high-order DG. Shocks, in particular, seed new  oscillations with time, because inside cells the smooth {\em inviscid} Euler equations are evolved -- in which there is no dissipation at all. Thus the entropy production required by shocks is simply not possible. Note that the oscillations are not only physically wrong, they can even cause negative density or pressure fluctuations in some cells, crashing the code.

One approach to prevent this are so-called slope limiters. In particular, the family of minmod slope limiters is highly successfully used in second-order finite volume methods. While use of them in DG methods is possible, applying them in  high order settings by discarding the high-order expansion coefficients whenever the slope limiter kicks in \citep[see][]{Schaal2015, guillet_high_order_2019} is defeating much of the effort to going to high order in the first place. Somehow constructing less aggressive high-order limiters that can avoid this is a topic that has seen much effort in the literature, but arguably only with still limited success. In fact, the problem of coping with shocks in high-order DG is fundamentally an issue that still awaits a compelling and reasonably simple solution. Recent advanced treatments had to resort to replacing troubled cells with finite volume solutions computed on small grid patches that are then blended with the DG solution \citep[e.g.][]{Zanotti2014, Markert:2021aa}.

We here return to the idea that this problem may actually be best addressed by resurrecting the  old idea of artificial viscosity \citep{persson_peraire_2006}.  In other inviscid hydrodynamical methods, in particular in the Lagrangian technique of smoothed particle hydrodynamics, it is evident and long accepted that artificial viscosity must be added to capture shocks. Because the conservation laws ultimately dictate the amount of entropy that needs to be created in shocks, the exact procedure for adding artificial viscosity is not overly critical. What is critical, however, is that the there is a channel for  dissipation and entropy production. It is also clear that shocks in DG can be captured in a sub-cell fashion only if the required dissipation is provided somehow, either through artificial viscosity that is ideally present only at the place of the shock front itself where it is really needed, or by literally capturing the shock by subjecting the ``troubled cell'' to a special procedure in which it is, for example, remapped to grid of finite volume cells.

\citet{persson_peraire_2006} suggested to use a discontinuity (or rather oscillation) sensor to detect the need for artificial viscosity in a given cell. For this, they proposed to  measure the relative contribution of the highest order Legendre basis functions in representing the state of the conserved fields in a cell. A solution of a smooth problem is expected to be dominated by the lower order weight coefficients, and statistically the low order weights should be much larger than their high order counterparts. In contrast, for highly oscillatory solutions in a cell (which often are created as pathological side-effects of discontinuities), the high order coefficients are more strongly expressed.

We adopt the same discontinuity sensor as \citet{persson_peraire_2006}. For every cell $K$, we can calculate the conserved variables $\boldsymbol{u}(\boldsymbol{x})$ using either the full basis in the normal way, 
\begin{equation*}
  \boldsymbol{u}(\boldsymbol{x})=\sum_{i=1}^{N(p)} \boldsymbol{w}^K_{l} \phi_{l}
\end{equation*}
or by omitting the highest order basis functions that are not present at the next lower expansion order, as
\begin{equation*}
  \boldsymbol{\hat{u}}(\boldsymbol{x})=\sum_{i=1}^{N(p-1)} \boldsymbol{w}^K_{l} \phi_{l}
\end{equation*}
The discontinuity/oscillatory sensor $S^{K}$ in cell $K$ can now be defined as
\begin{equation}
\label{eq:discontinuity_sensor1}
    S^{K}=\frac{\int_K (u-\hat{u})(u-\hat{u}) {\rm d}V }{\int_K u(x) u(x) \,{\rm d}V },
\end{equation}
where we restrict ourselves to one component of the state vector, the density field. Note that due to the orthogonality of our basis functions, this can be readily evaluated as 
\begin{equation}
\label{eq:discontinuity_sensor2}
    S^{K}=\frac{\sum\textbf{}_{l=N_{p-1}}^{N_p} [w^K_l]^2}{\sum_{l=1}^{N_p} [w^K_l]^2 }
\end{equation}
in terms of sums over the squared expansion coefficients. While we have $0 \le S^{K} \le 1$, we expect $S^{K}$ to generally assume relatively small values even if significant oscillatory behaviour is already present in $K$, simply because the natural magnitude of the expansion coefficients declines with their order rapidly.  \citet{persson_peraire_2006} argue that the coefficients should scale as $1/p^2$ in analogy with the scaling of Fourier coefficients in 1D, so that typical values for $S^{K}$ in case oscillatory solutions are present may scale as $1/p^4$. Our tests indicate a somewhat weaker scaling dependence, however, for oscillatory solutions developing for identical ICs, where the troubled cells scale approximately as $S^K \sim 1/p^2$ as a function of order.

In the approach of \citet{persson_peraire_2006}, artificial viscosity is invoked in cells once their $S^K$ value exceeds a threshold value, above which it is ramped up smoothly as a function of $S^K$ to a predefined maximum value. While this approach shows some success in controlling shocks in DG, it is problematic that strong oscillations need to be present in the first place {\em before} the artificial viscosity is injected to damp them. In a sense, some damage must have already happened before the fix is applied. 

For capturing shocks we therefore argue it makes more sense to resort to a physical shock sensor which detects rapid, non-adiabatic compressions in which dissipation should occur. We therefore propose here to adapt ideas widely used in the SPH literature \citep{Morris1997, Cullen2010}, namely to consider a time-dependent artificial viscosity field that is integrated in time using suitable source and sink functions. Adopting a dimensionless viscosity strength $\alpha(\boldsymbol{x}, t)$, we propose the evolutionary equation
\begin{equation}
    \frac{\partial \alpha}{\partial t} = \dot\alpha_{\rm shock} + \dot\alpha_{\rm wiggles}   -\frac{\alpha}{\tau}
    \label{eqn:timedependentviscosity}
\end{equation}
for steering the spatially and temporarily variable viscosity. For the moment we use a simple shock sensor $\dot\alpha_{\rm shock} = f_v \max(0, -\boldsymbol{\nabla}\cdot \boldsymbol{v})$ based on detecting compression, where $f_v\sim 1.0$ can be modified to influence how rapidly the viscosity should increase upon strong compression. In the absence of sources, the viscosity decays exponentially  on a timescale
\begin{equation}
\tau = f_\tau \frac{h}{p\,c_s},
\end{equation}
where $h/p$ is the expected effective spatial resolution at order $p$, $c_s$ is the local sound speed, and $f_\tau \sim 0.5$ is a user-controlled parameter for setting how rapidly the viscosity decays again after a shock transition.

\begin{figure}
	\includegraphics[width=88mm]{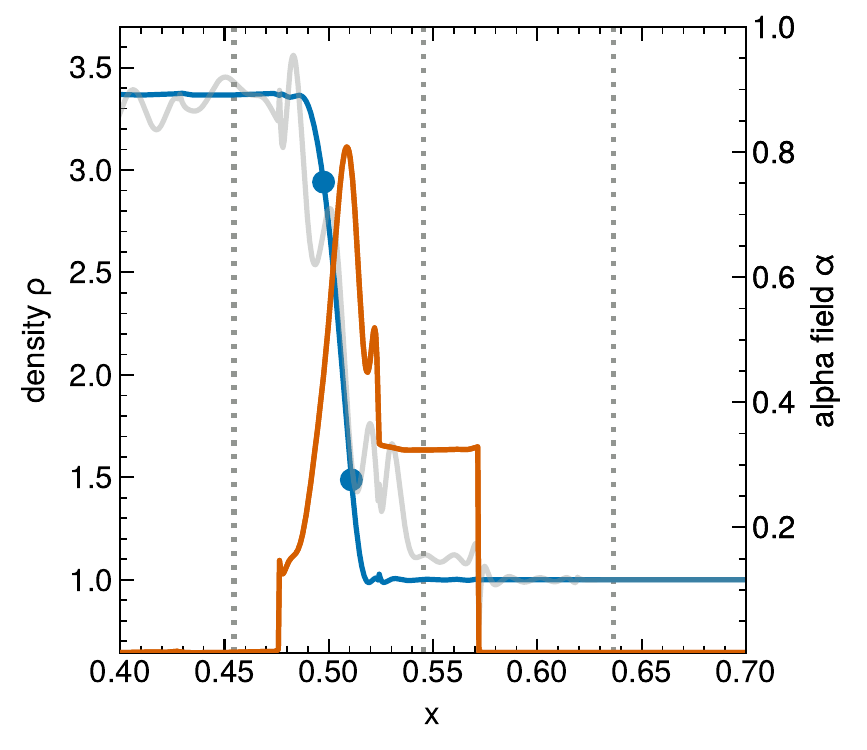}\vspace*{-0.4cm}
    \caption{Zoom into a Mach number ${\cal M}=4$ shock that is simulated with order $p=9$. The upstream gas has unit density and unit pressure. Individual mesh cell boundaries are indicated with dotted lines. The density field obtained with artificial viscosity included is shown as a solid blue, while the result without artificial viscosity is shown as a grey line in the background. The artificial viscosity field itself is shown as orange line (scale on the right). The analytic shock position at the displayed time is at $x=0.5$, in the middle of one of the mesh cells. The circles mark the locations where the density has reached 20 and 80 percent, respectively, of the shock's density jump. We use the distance $\Delta x_\textrm{shock}$ of the corresponding points as a measure of the shock width.}
    \label{fig:shockzoom_viscosity}
\end{figure}

\begin{figure}
	\includegraphics[width=\columnwidth]{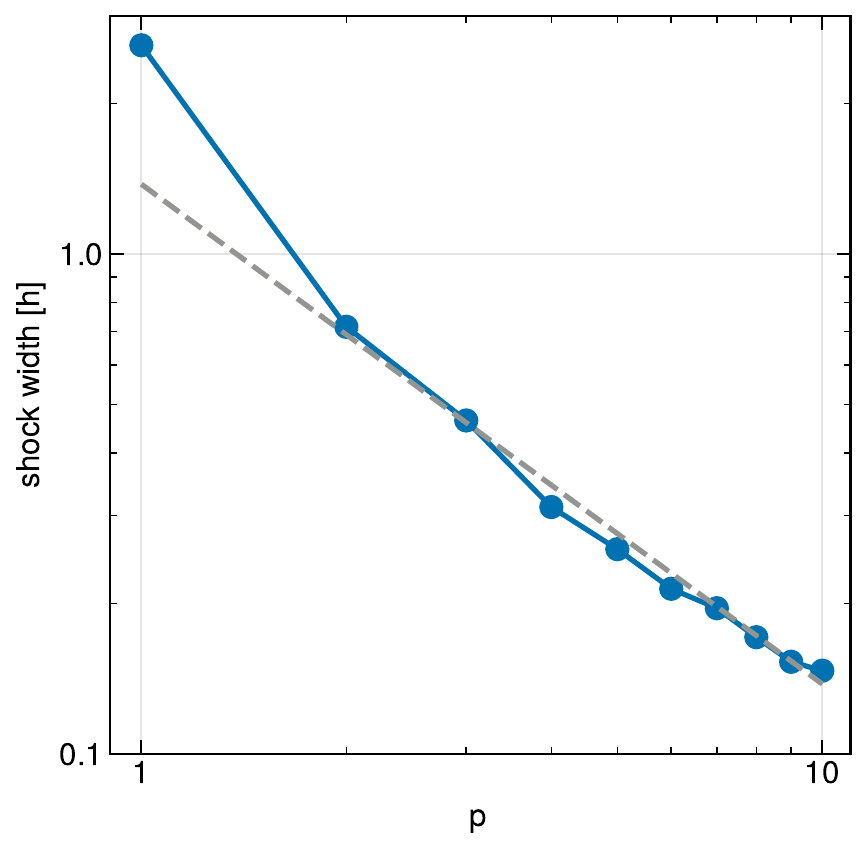}\vspace*{-0.4cm}
    \caption{Shock width in units of the cell size as a function of the order $p$ of our DG code, for a Mach number ${\cal M}=4$ shock that runs into gas at rest. The dashed line marks a $\Delta x_\textrm{shock} \propto 1/p$ power law, which accurately describes our measurements, except for the lowest order result with piece-wise constant states, which is so highly diffusive that it does not require any artificial viscosity. }
    \label{fig:shockwidth_viscosity}
\end{figure}

Finally, the term $\dot\alpha_{\rm wiggles}$ in Equation~(\ref{eqn:timedependentviscosity}) is a further source term added to address the occurrence of oscillatory behaviour away from shocks. In fact, this typically is seeded directly ahead of strong shocks, for example when the high-order polynomials in a cell with a  shock trigger oscillations in the DG cell directly ahead of the shock through coupling at the interface. Another typical situation where oscillations can occur are sharp, moving contact discontinuities. Here the shock sensor would not be effective in supplying the needed viscosity as there is no shock in the first place. We address this problem by considering the {\em rate of change} of the oscillatory senor $S^K$ as a source for viscosity, in the form
\begin{equation}
\dot\alpha_{\rm wiggles}  = f_w \max\left(0, \frac{{\rm d} \log S^K}{{\rm d}t}  \right), \\
\end{equation}
for $S^K > S_{\rm onset}$, otherwise $\dot\alpha_{\rm wiggles} = 0$.
When ${\rm d} \log S^K/{{\rm d}t}$ is positive and large, oscillatory behaviour is about to grow and the cell is on its way to become a troubled cell, indicating that this should better be prevented with local viscosity. In this way, oscillatory solutions can be much more effectively controlled than waiting until they already reached a substantial size. It is nevertheless prudent to restrict the action of this viscosity trigger to cells that have $S_K$ above a minimum value $S_\textrm{onset}$, otherwise the code would try to suppress even tiny wiggles, which would invariably lead to very viscous behaviour. In practice, we set $S_{\rm onset} = 10^{-4}/p^2$, and we compute ${\rm d} \log S^K/{{\rm d}t}$ based on the time derivatives of the weights of the previous timestep.

We add $\alpha$ as a further field component to our state vector $\boldsymbol{u}$, meaning that it is spatially variable and is expanded in our set of basis functions. We do not advect the $\alpha$ field with the local flow velocity as to allow it to fall behind moving discontinuities and to fully suppress any excited oscillations there. Also, advecting the  $\alpha$ field at high order would require a limiting scheme for this field itself. Note that in the post-/pre-shock region we can assume the first term of Eq.~(\ref{eqn:timedependentviscosity}) to be unimportant. Once the wiggles are suppressed the second term disappears as well, so that then the default choice of parameters suppresses any existing $\alpha$ field to percent level in a handful of time steps. Only the shock sensor source function is actually variable in a cell, whereas our oscillatory sensor affects the viscosity throughout a cell.

Finally, the actual viscosity applied in the viscous flux of Eqn.~(\ref{eq:navier-stokes-basic}) is parameterized as
\begin{equation}
    \epsilon = \alpha c_s \frac{h}{p}, 
    \label{eq:viscosu_flux_with_factor}
\end{equation}
and we impose a maximum allowed value of $\alpha_{\rm max} = 1$, primarily as a means to prevent overstepping and making the scheme violate the the von Neumann stability requirement for explicit integration of the diffusion equation, which would cause immediate numerical instability. Since our timestep obeys the Courant condition, this is fortunately not implying a significant restriction for effectively applying the artificial viscosity scheme, but it imposes an upper bound that can be used safely without making the time-integration unstable.

We have found that the above parameterisation works quite reliably, injecting viscosity only at discontinuities and when spurious oscillations need to be suppressed, while at the same time not smoothing out solutions excessively. Figure~\ref{fig:shockzoom_viscosity} shows an example for a Mach number $\mathcal{M} = 3$ shock that is incident from the left on gas with unit density and unity pressure, and adiabatic index $\gamma=5/3$. The simulation has been computed at order $p=9$, and at the displayed time, the shock position should be at $x=0.5$, for a mesh resolution of $h=1.0/21$. We show our DG result as a thick blue line, and also give the viscosity field $\alpha(x)$ as a red line. Clearly, the shock is captured at a fraction of the cell size, with negligible ringing in the pre- and post-shock regions. This is achieved thanks to the artificial viscosity, which peaks close to the shock center, augmented by additional weaker viscosity in the cell ahead of the shock, which would otherwise show significant oscillations as well. This becomes clear when looking at the solution without artificial viscosity, which is included as a grey line in the background. 

The blue circles in Fig.~\ref{fig:shockzoom_viscosity} mark the places in which the solution has reached 20 and 80 percent of the height of the shock's density jump. We can operationally define the difference in the corresponding $x$-coordinates as the width $\Delta x_\textrm{shock}$ with which the shock is numerically resolved. In Figure~\ref{fig:shockwidth_viscosity} we show  measurements of the shock width for the same set-up, except for varying the employed order $p$. We see that the shock width declines with higher order, accurately following the desired relationship $\Delta x_\textrm{shock} \propto 1/p$, except for the lowest order $p=1$, which deviates towards  broader width compared to the general trend. The importance of this result for the DG approach can hardly be overstated, given that it has been a nagging problem for decades to reliably capture shocks at sub-cell resolution in DG without having to throw away much of the higher resolution information. The result of Figure~\ref{fig:shockwidth_viscosity} essentially implies that shocks are resolved with the same width for a fixed number of degrees of freedom, independent of the employed order $p$. Whereas using higher order at a fixed number of degrees of freedom is thus not providing much of an advantage for making shocks thinner compared to using more cells, it at least does not degrade the solution. But smooth parts of a solution can then still benefit from the use of higher order.

In total our artificial viscosity method uses five parameters, one for each of the three terms of Eq.~(\ref{eqn:timedependentviscosity}), a further general scaling factor $\alpha$ which is applied to the total viscous flux as defined in Eq.~(\ref{eq:viscosu_flux_with_factor}), and an onset threshold  $S_{\rm onset}$. In this way we are able to individually control the suppression of shocks, wiggles and the decay time of the viscous field as well as the total magnitude of viscous flux. The default values we adopted for these parameters throughout this work are 
     $\alpha = 1.0$, 
     $f_v = 2.5$, 
     $f_\tau = 0.5$, 
     $f_w = 0.2$, and
     $S_{\rm onset} = 10^{-4}$.

\begin{figure*}
	\resizebox{18cm}{!}{\includegraphics{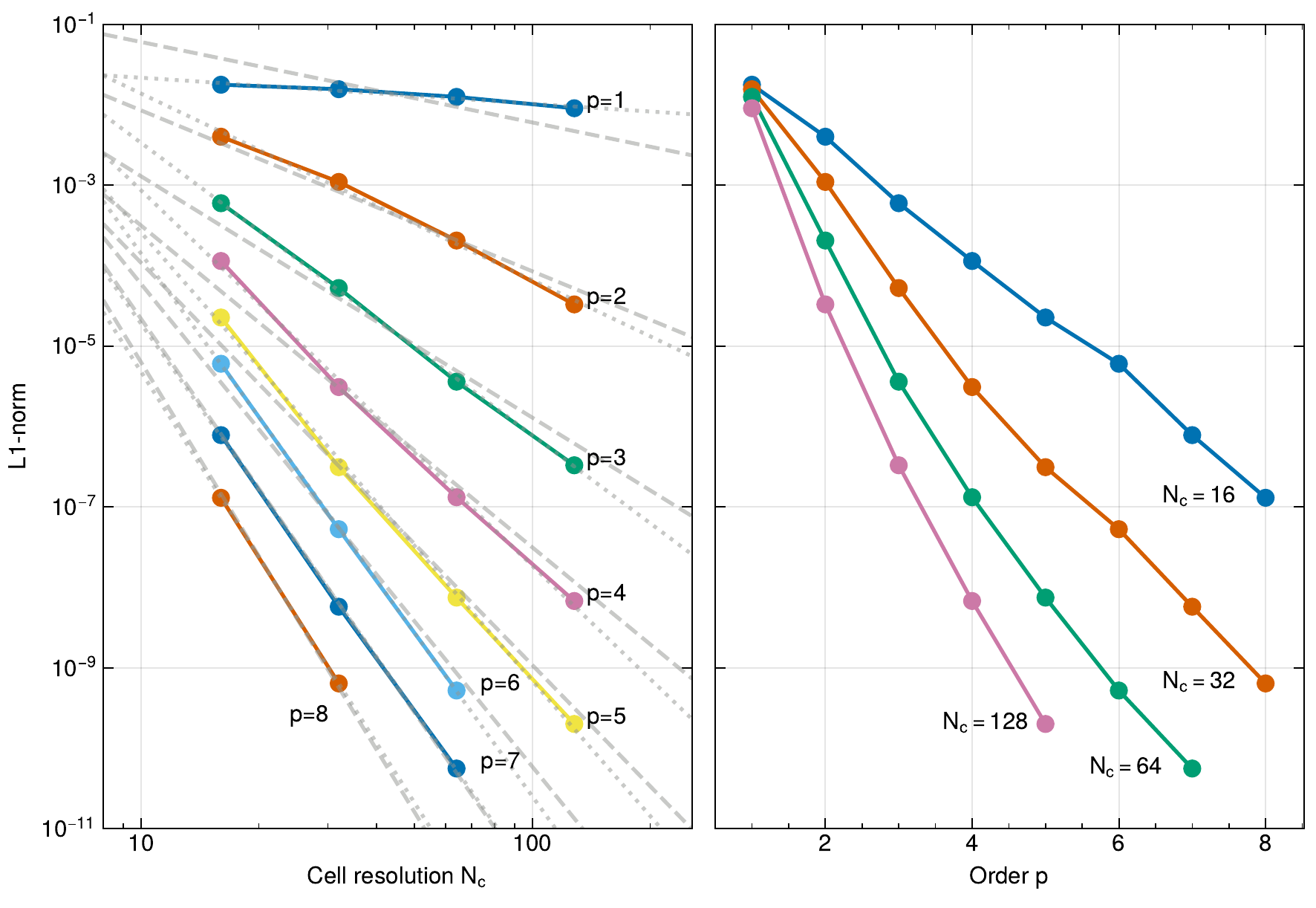}}%
    \caption{Convergence of the \citet{Yee1999, Yee2000} vortex when evolved for $t=10.0$ time units. The left panel shows the error norm in the density fields as a function of spatial grid resolution, for 8 different orders $p$ of our DG scheme. The measured convergence orders for $L1$ (colored lines) are close to the expected $L1 \propto N_{\rm c}^{-p}$ power-laws (dashed grey lines). The actually achieved convergence orders (fitted power-laws, shown as dotted lines) are typically even slightly better than expected, except for the lowest order $p=1$. The panel on the right-hand side shows the same data, but  as a function of DG order $p$, using a log-linear plot. For fixed grid resolution, the error declines {\em exponentially} with the order $p$ of the scheme, highlighting the very fast improvement of accuracy when the DG order is increased. We note that the imposed periodic boundaries for the chosen box size of 10 lead to an edge effect which puts the lower boundary of the L1-norm to $\sim 10^{-11}$.}
    \label{fig:isentropic_vortex_convergence}
\end{figure*}

\subsection{Positivity limiter}
\label{sec:positivity_limiter}

With our artificial viscosity approach described above we intend to introduce the necessary numerical viscosity where needed, such that slope limiting becomes obsolete. However, for further increasing robustness of our code, it is desirable that it also runs stably if a too weak or no artificial viscosity is specified, or if its strength is perhaps locally not sufficient for some reason in a particularly challenging flow situation. To prevent a breakdown of the time evolution in this case, we consider an optional positivity limiter following \citet{Zhang2010} and \citet{Schaal2015}. This can be viewed as a kind of last line of defense against the occurrence of oscillations in a solution that ventures into the regime of unphysical values, such as negative density or pressure. The latter can happen even for arbitrarily small timesteps, especially when higher order methods are used where such robustness problems tend to be more acute. 

Finite-element and finite-volume hydrodynamical codes typically employ procedures such as slope limiters to cope with these situations, this means they locally reduce the order of the scheme (effectively making it more diffusive) by discarding high-order information. A similar approach is followed by the  positivity limiter described here, which is based on \cite{Schaal2015}, with an important difference in how we select the evaluation points. We stress however that the positivity limiter is not designed to prevent oscillations, only to reduce them to a point that still allows the calculation to proceed.

For a given cell, we first determine the average density $\overline{\rho}$ in the cell, which  is simply given by the 0-th order expansion coefficient for the density field of the given cell, and we likewise determine the average pressure $\overline{p}$ of the cell. If either $\overline{\rho}$ or $\overline{p}$ is negative, a code crash is unavoidable. 

Otherwise, we define a lowest permissible density $\rho_\textrm{bottom} =  10^{-6} \bar{\rho}$.   Next, we consider the full set of quadrature evaluation points $\{\boldsymbol{x}_i\}$ relevant for the cell, which is the union of the points used for internal volume integrations and the points used for surface integrals on the outer boundaries of the cell. We then determine the minimum  density $\rho_{\rm min}$ occurring for the field expansions among these points. In case $\rho_\textrm{min} < \rho_\textrm{bottom}$, which includes the possibility that $\rho_{\rm min}$ is negative, we calculate a reduction factor $f = \left( \bar{\rho} - \rho_\textrm{bottom} \right) / \left( \bar{\rho} - \rho_\textrm{min} \right)$  and replace all higher order weights of the cell with
\begin{equation}
 \boldsymbol{w}'^K_{l} = f\, \boldsymbol{w}^K_{l}  \;\;\mbox{for}\;\;l > 1.
\end{equation}  
This limits the minimum density appearing in any of the discrete calculations to $\rho_\textrm{bottom}$. By applying the correction factor $f$ to all fields and not just the density, we avoid to potentially amplify relative fluctuations in the velocity and pressure fields. 

We proceed similarly for limiting pressure oscillations, except that here no simple reduction factor can be computed to ensure that $p_\textrm{min}$ stays above $p_\textrm{bottom}$, due to the non-linear dependence of the pressure on the energy, momentum and density fields. Instead, we simply adopt $f=0.5$ and repeatedly apply the pressure limiter until $p_{\rm min} \ge p_\textrm{bottom}$.

In our test simulations the positivity limiter, as expected, does not trigger for inherently smooth problems and thus is in principle not needed. However, when starting simulations with significant discontinuities in the initial conditions, the positivity limiter usually kicks in at the start for a couple of timesteps, especially for high order simulations, until the artificial viscosity is able to tame the spurious oscillations, making the positivity limiter superfluous in the subsequent evolution.

\section{Basic tests}
\label{SecBasicTests}

In this section we consider a set of basic tests problems that establish the accuracy of our new code both for smooth problems, as well as for problems containing strong discontinuities such as shocks or contact discontinuities. We shall begin with a smooth hydrodynamic problem that is suitable for verifying code accuracy for the inviscid Euler equations. We then turn to testing the diffusion solver of the code, as an indirect means to test the ability of our approach to stably and accurately solve the viscous diffusion inherent in the  Navier-Stokes equations. We then consider shocks and the supersonic advection of a discontinuous top-hat profile to verify the stability of our high-order approach when dealing with such flow features. Applications to Kelvin-Helmholtz instabilities and driven turbulence are treated in separate sections.

\subsection{Isentropic vortex}

The  isentropic  vortex problem of \citet{Yee1999, Yee2000} is a time-independent smooth vortex flow, making it a particularly useful test for the accuracy of higher-order methods, because they should reach their theoretically optimal spatial convergence order if everything is working well \citep[e.g.][]{Schaal2015, Pakmor2016}. We follow here the original setup used in \citet{Yee1999} by employing a domain with extension $[-5, 5]^2$ in 2D and an initial state given by:
\begin{eqnarray} \label{eq:yee_vortex_vx}
v_x(\boldsymbol{r}) & = & -\frac{\beta y}{2\pi} \exp\left(\frac{1-r^2}{2}\right) \\
v_y(\boldsymbol{r}) & = & \frac{\beta x}{2\pi} \exp\left(\frac{1-r^2}{2}\right)\\
u(\boldsymbol{r}) & = & 1 - \frac{\beta^2 }{8\gamma \pi^2} \exp\left({1-r^2}\right) \\
\rho(\boldsymbol{r}) & = & [ (\gamma-1) u(\boldsymbol{r}) ]^{\frac{1}{\gamma-1}}
\end{eqnarray}
where we choose $\gamma = 1.4$, and $\beta =5 $. We evolve the vortex with different DG expansion order $n$ and different mesh resolutions $N_{\rm grid}^2$ until time $t=10$, and then measure the resulting L1 approximation error of the numerical result for the density field relative to the analytic solution (which is identical to the initial conditions). In order to make the actual measurement of $L1$ independent of discretization effects, we use $n+2$ Gaussian quadrature for evaluating the volume integral appearing in Eq.~(\ref{eq:l1_norm}). Likewise, we use this elevated order when projecting the initial conditions onto the discrete realization of DG weights of our mesh.

In Figure~\ref{fig:isentropic_vortex_convergence} we show measurements of the L1 error as a function of grid resolution $N_{\rm grid}$, for different expansion order from $p=1$ to $p=8$. The left panel shows that the errors decrease as power laws with spatial resolution for fixed $n$, closely following the expected convergence order $L1\propto N_{\rm grid}^{-p}$ in all cases (except for the $p=1$ resolution, which exhibits  slightly worse behavior -- but this order is never used in practice because of its dismal convergence properties). 

Interestingly, the data also shows that for a given grid resolution, the L1 error goes down {\em exponentially} with the order of the scheme. This is shown in the right panel of Fig.~\ref{fig:isentropic_vortex_convergence}, which shows the L1 error in a log-linear plot as a function of order $p$, so that exponential convergence manifests in a straight decline.  This particularly rapid decline of the error with $p$ for smooth problems makes it intuitively clear that it can be advantageous to go to higher resolution if the problem at hand is free of true physical discontinuities.

\subsection{Diffusion of a Gaussian pulse}

To test our procedures for simulating the diffusion part of our equations, in particular our treatment for estimating surface gradients at interfaces of cells, we first consider the diffusion of a Gaussian pulse, with otherwise stationary gas properties. For simplicity, we consider gas at rest and with uniform density and pressure, and we consider the evolution of a small Gaussian concentration of a passive tracer dye under the action of a constant diffusivity.

For definiteness, we consider a tracer concentration $c(\vec{x}, t)$ given by 
\begin{equation}
c(\vec{x}, t) = c_b + \sum_{\vec j} \frac{c_g}{2\pi \sigma^2} \exp\left(-\frac{(\vec{x} - \vec{j})^2}{2\sigma^2}\right),
\;\;\;\mbox{with}\; \sigma^2 = 2\eta t,
\label{eqn:diffusion_spike}
\end{equation}
placed in a unit domain $[-0.5, 0.5]^2 $ in 2D with periodic boundary conditions. Here the sum over $\vec{j}$ effectively accounts for a Cartesian grid of Gaussian pulses spaced one box size apart in all dimensions to properly take care of the periodic boundary conditions. If we adopt a fixed diffusivity $\eta$ and initialize $c(\vec{x}, t)$ at some time $t_0$, then the analytic solution of equation~(\ref{eqn:passivetracer}) tells us that eqn.~(\ref{eqn:diffusion_spike}) will also describe the dye concentration at all subsequent times $t > t_0$.

\begin{figure}
	\includegraphics[width=88mm]{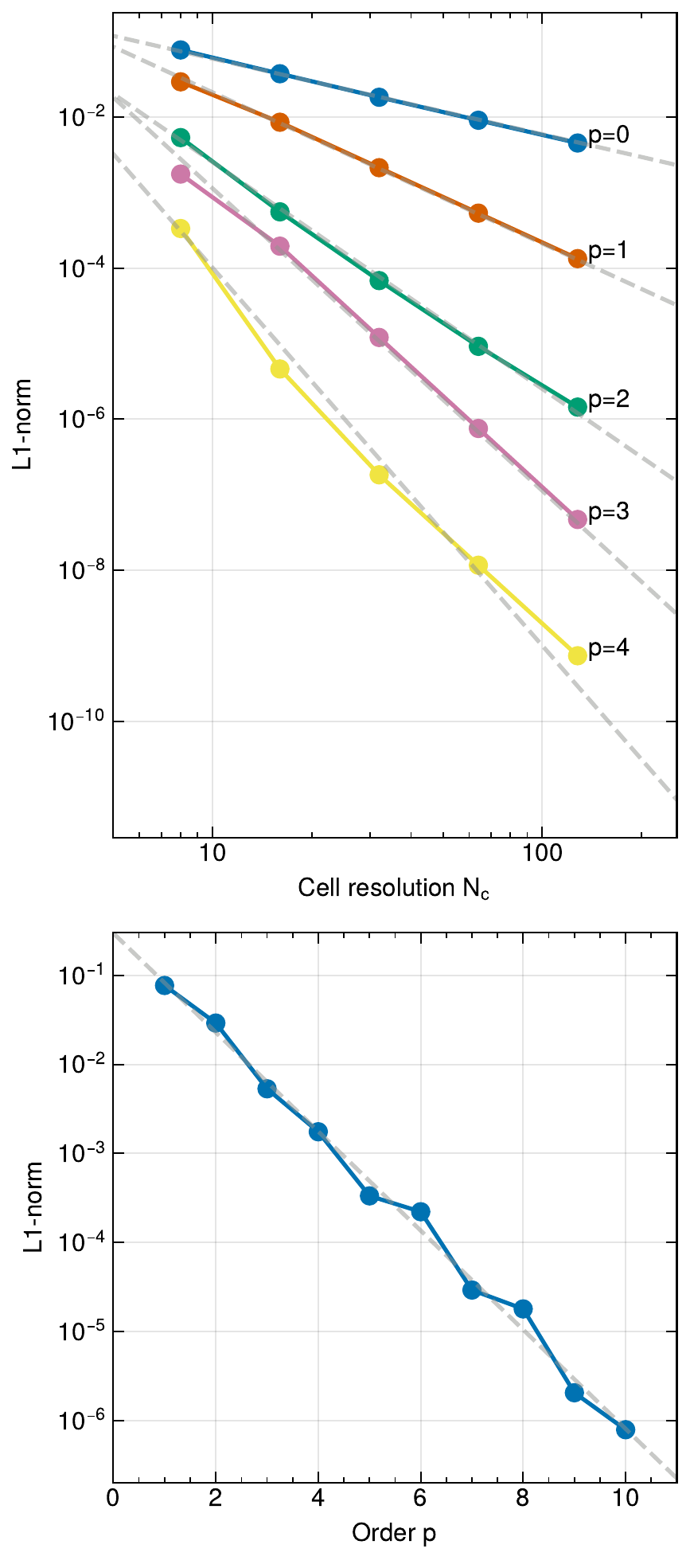} 
    \caption{Convergence of the diffusion process of a Gaussian profile when started from a smooth state. The top panel shows results for runs carried out at different mesh resolution $N_{\rm c}$ and DG expansion order $p$, as labelled. For fixed expansion order, the L1 error declines as a power law as a function of the spatial grid resolution, with the slope of the the expected  convergence rate. In the bottom panel, we show the error as a function of order at a fixed grid resolution of $N_{\rm c} = 8$. In this case, the error declines exponentially as a function of the expansion order. }
    \label{fig:gaussian_diffusion}
\end{figure}

For definiteness, we choose $\eta = 1/128$, $c_b = 1/10$, $c_g=1$, and $t_0=1$, and examine the numerically obtained results at time $t = t_0 + 3 = 4$ by computing their L1 error norm with respect to the analytic solution. In the top panel of Fig.~\ref{fig:gaussian_diffusion}, we show the convergence of this diffusion process as a function of the number of grid cells used, for the first five DG expansion orders. Reassuringly, the L1 error norm decays as a power-law with the cell size, in each case with the expected theoretical optimum $L1 \propto N_{\rm cells}^{-p}$. This shows that our treatment of the surface derivatives is not only stable and robust, but is also able to deliver high-order convergence. 

The bottom panel of Figure~\ref{fig:gaussian_diffusion} shows that this also manifests itself in an exponential convergence as function of DG expansion order when the mesh resolution is kept fixed. For this result, we adopted $N_{\rm cells}=8$ and went all the way to 10-th order.

While these results do not directly prove that our implementation is able to solve the full Navier-Stokes equations at high-order, they represent an encouraging prerequisite. Also, we note that both the version without viscous source terms (i.e.~the Euler equations), as well as the viscous term itself when treated in isolation converges at high order. We will later on compare to a literature result for the Kelvin-Helmholtz instability in a fully viscous simulation to back up this further and to test a situation where the full Navier-Stokes equations are used.

\begin{figure}
\begin{center}
	\includegraphics[width=70mm]{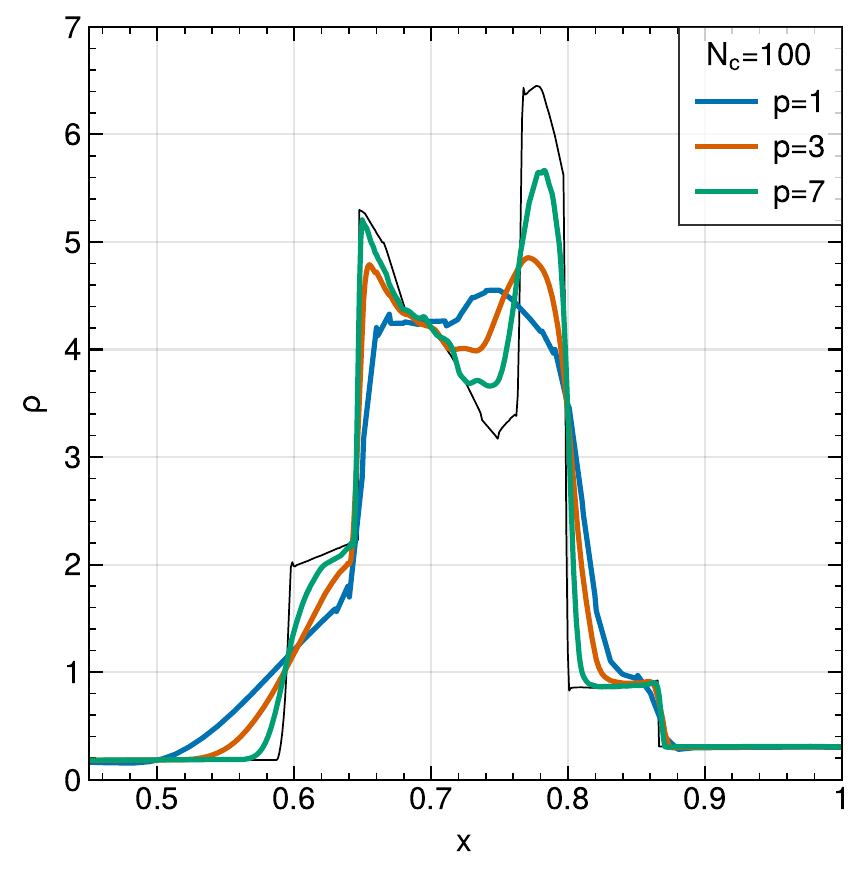}
	\end{center}
    \caption{Double blast wave problem at fixed spatial resolution, but for increasing DG order. This shows clearly that our new artificial viscosity method can cope with strong shocks, and that adding higher order information is still worthwhile in treating problems with strongly interacting shocks. For reference, a high resolution result with $N_\textrm{c}=10000$, $p=1$ is shown as thin black line.}
    \label{fig:double_blast_wave}
\end{figure}

\begin{figure}
	\includegraphics[width=88mm]{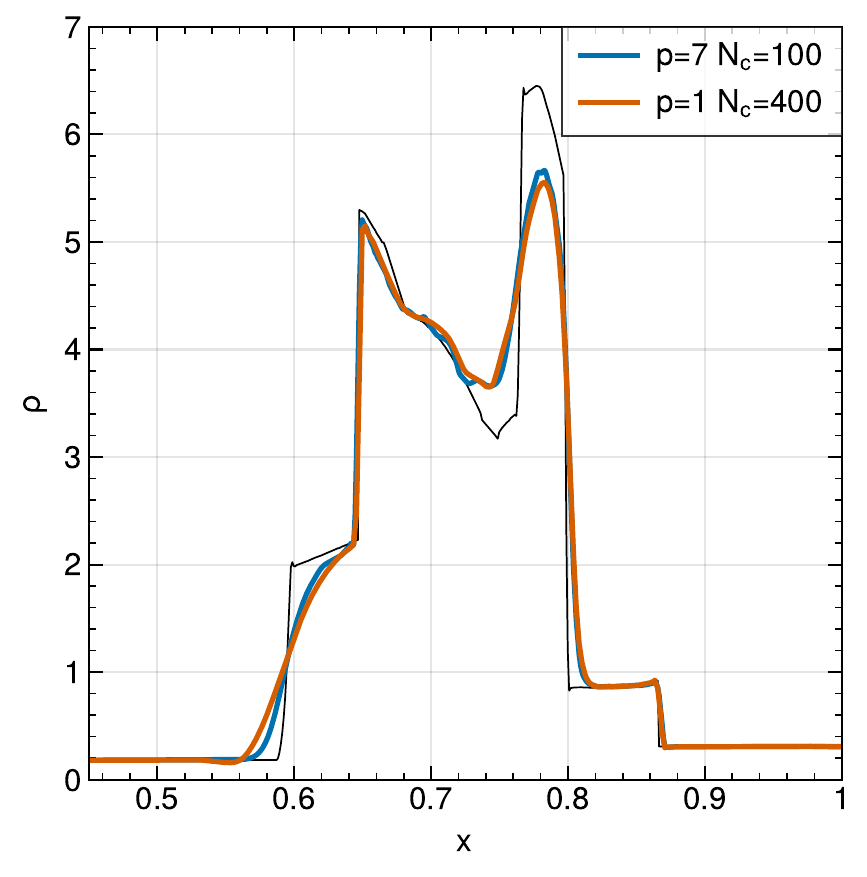}
    \caption{Double blast wave problem at fixed number of degrees of freedom for two different combinations of order and spatial resolution. This shows that for strong shocks the total number of degrees of freedom determines accuracy of our solution. For reference, a high resolution result with $N_\textrm{c}=10000$, $p=1$ is shown as thin black line.}
    \label{fig:doubleblastwave_k1_nc_200_vs_800}
\end{figure}

\begin{figure}
\begin{center}
	\resizebox{8cm}{!}{\includegraphics{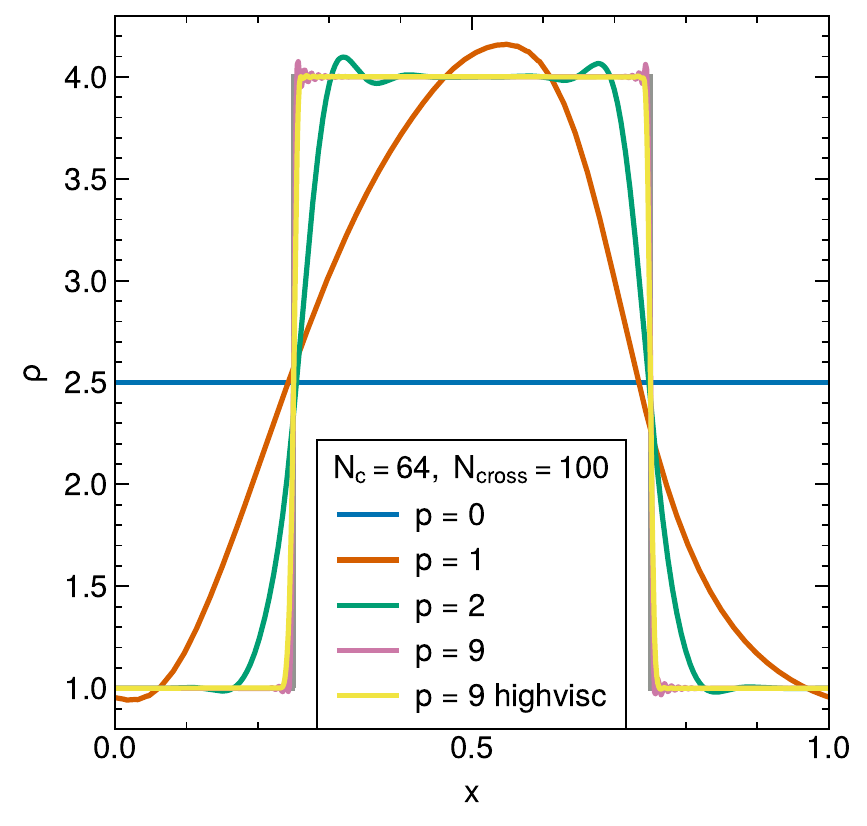}}
	\resizebox{8cm}{!}{\includegraphics{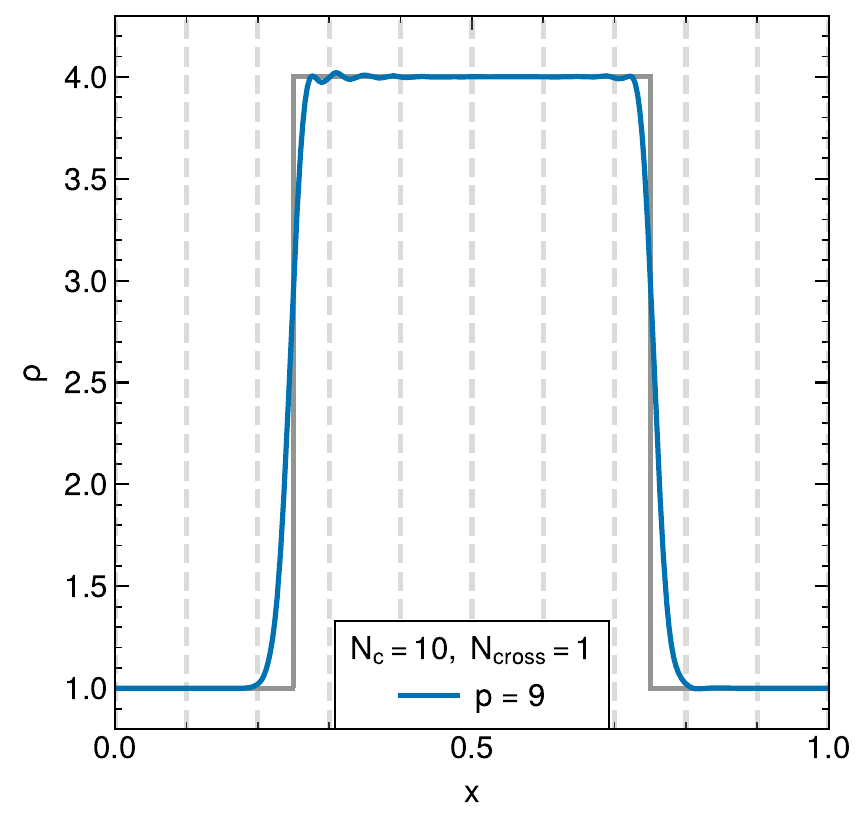}}
	\end{center}
    \caption{ \textit{Top panel:} Square advection problem at $t=1.0$ for different expansion orders $p$ using 64 grid cells in each case. At this time, the top hat profile has been advected 100 times through the box. The initial profile, which is the analytic solution in this case, is shown as a solid grey line in the background. Different numerical results are given for polynomial orders $p=0, 1, 2$, and $p=9$, as well as for $p=9$ with a higher artificial viscosity setting for stronger wiggle suppression. \textit{Bottom panel:} Square advection problem at $t=0.01$ for $p=9$ using 10 grid cells. The profile has been advected through the box once. The dotted vertical lines indicate grid cell borders. Sub-cell shock capturing can be observed.
    }
    \label{fig:square_advection_shape}
\end{figure}

\begin{figure}
\begin{center}
	\resizebox{8cm}{!}{\includegraphics{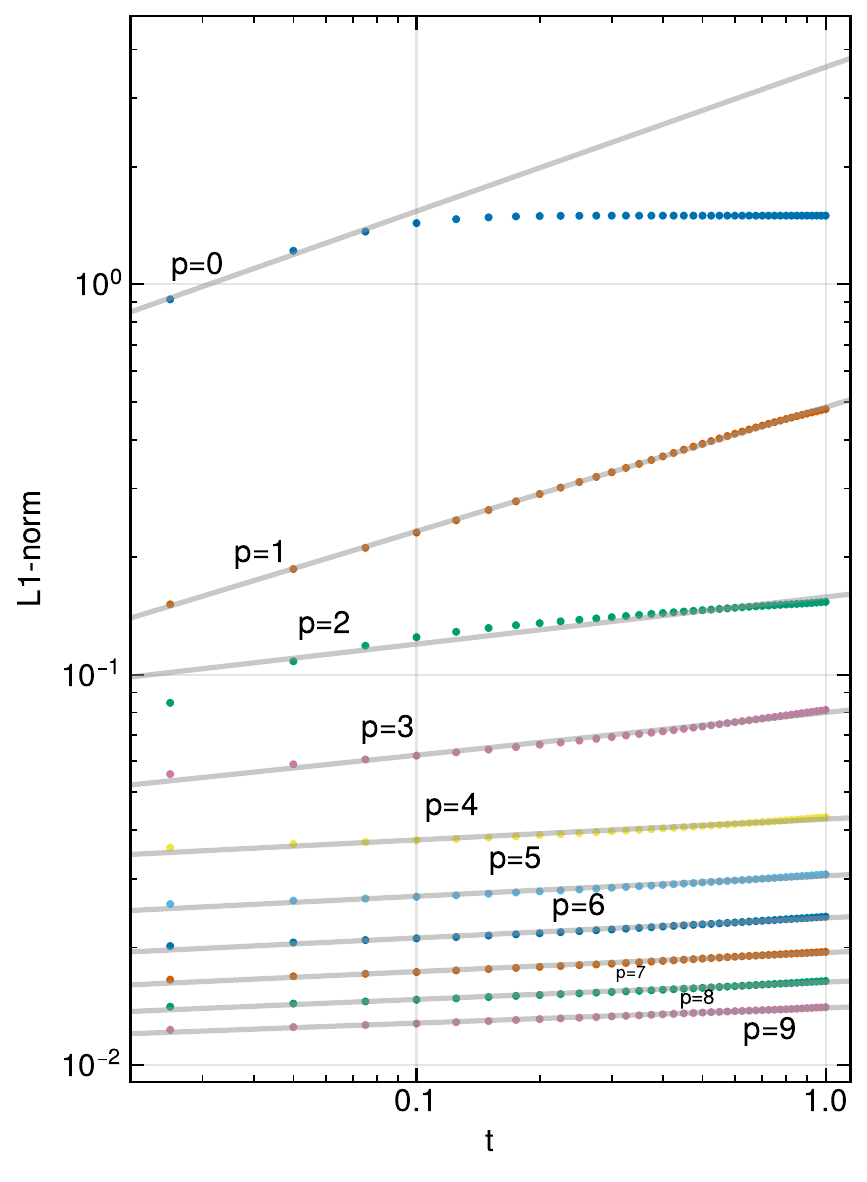}}
	\end{center}
    \caption{Time evolution of the L1 error norm for the density in the square advection problem, calculated for polynomial orders $p=0$ to $p=9$ (from top to bottom) using 64 grid cells in each case. The individual measurements for numerical outputs are shown with filled circles, the lines are power-law fits $L1 \propto t^n$. Note that not only the absolute error at any given time declines with increasing order $p$, also the slopes $n$ become progressively shallower. This means that the numerical diffusivity of the code becomes smaller for higher order, reducing advection errors substantially. The measured slopes $n$ for the $p=0$ to $p=9$ cases are in that sequence: 0.427, 0.335, 0.172, 0.056, 0.054, 0.049, 0.048, 0.046, 0.039, and 0.028. In the $p=0$ case, only the first three points were used to measure the slope.
    }
    \label{fig:square_advection_L1}
\end{figure}

\subsection{Double blast wave}

To test the ability of our DG approach to cope with strong shocks, particularly at high order, we look at the classic double blast wave problem of \citet{Woodward1984}.  The initial conditions are defined in the one-dimensional domain $[0,1]$  for a gas of unit density and adiabatic index $\gamma = 7/5$, which is initially at rest. By prescribing two regions of very high pressure,  $P=1000$ for  $x < 0.1 $, and $P=100$ for $x > 0.9$, in an otherwise low-pressure $P=0.1$ background, the time evolution is characterized by the launching of very strong shock and rarefaction waves that collide and interact in complicated ways. Because of the difficulty of this test for shock-capturing approaches, it has often been studied in previous work to examine code accuracy and robustness \citep[e.g.][]{Stone2008, Springel2010}.

In order to highlight differences due to different DG orders, we have run deliberately low-resolution realizations of the problem, using 100 cells of equal size within the region $[0,1]$. We have then evolved the initial conditions with order$p=2$, $p=4$, or $p=8$. Furthermore, we examine a run done with four times as many cells carried out at order $p=2$. This latter simulation has the same number of degrees of freedom as the $p=8$ simulation, and thus should have a similar effective spatial resolution.  For comparison purposes, we use a simulation carried out with 10000 cells at order $p=2$, which can be taken as a result close to a converged solution. All simulations were run with our artificial viscosity implementation using our default settings for the method (which do not depend on order $p$).

In Figure~\ref{fig:double_blast_wave}, we show the density profile at the time $t = 0.038$, as done in many previous works, based on our 100 cell runs. Clearly, the shock fronts and contact discontinuities of the problem are quite heavily smoothed out for the $p=2$ run with 100 cells, due to the low resolution of this setup. However, the quality of the result can be progressively improved by going to higher order while keeping the number of cells fixed, as seen by the results for $p=4$ and $p=8$. This is in itself important. It shows that even problems dominated by very strong physical discontinuities are better treated by our code when higher order is used. The additional information this brings is not eliminated by slope-limiting in our approach, thanks to the sub-cell shock capturing allowed by our artificial viscosity technique. 

Finally, in Figure~\ref{fig:doubleblastwave_k1_nc_200_vs_800} we compare the $p=7$, 100 cell result to the $p=1$ result using 400 cells. Recall that the order of the method is $p+1$ and the total number of degrees of freedom of the two simulations is therefore the same. We find essentially the same quality of the results, which is another important finding. This demonstrates that to first order only the number of degrees of freedom per dimension is important for determining the ability of our DG code to resolve shocks. Putting degrees of freedom into higher expansion order instead of into a larger number of cells is thus not problematic for representing shocks. At the same, it also does not bring a clear advantage for such flow structures. This is because shocks are ultimately always broadened to at least the spatial resolution limit. Real discontinuities therefore only converge with 1st order in  spatial resolution, and high-order DG schemes do not provide a magic solution for this limitation as their effective resolution is set by the degrees of freedom. Still, as our results show, DG can be straightforwardly applied to problems with strong shocks using our artificial viscosity formulation. When there is a mixture of smooth regions and shocks in a flow, the smooth parts can still benefit from the higher order accuracy while the shocks are rendered with approximately the same accuracy as done with a second-order method with the same number of degrees of freedom.

It is interesting to compare our results to other results in the literature. First we compare to an older DG implementation with a moment limiter by \citet{krivodonova2007}. At low orders their mode by mode limiter performs marginally better than our implementation, but as it was pointed by   \citet{vilar2019} the mode by mode limiting does not work well for higher orders. In contrast, our method remains stable and offers steadily improving results as the order is increased. \citet{vilar2019} uses a DG implementation with \textit{a posteriori} limiting where troubled cells are detected at the end of the time-step and then recomputed using a finite-volume method and flux reconstruction. Their approach is also able to resolve the complicated interactions of shocks and rarefaction waves and yields a steadily improving result with higher resolution as well. To demonstrate that our DG implementation is competitive with state-of-art weighted essentially non-oscillatory (WENO) schemes we compare our results with those reported by \citet{zhao2017}. They simulated this problem using an 8-th order WENO scheme with 400 grid cells. The WENO implementation performs here somewhat better at a given number of degrees of freedom compared to our DG method. Note that the number of degrees of freedom in a WENO scheme is order independent and therefore our $p=7$ run at at 100 cells has twice the number of degrees of freedom as their 8-th order run with 400 cells, yet their result is closer to the ground truth than ours.

\subsection{Advection of a top-hat pulse}

Next we consider the problem of super-sonically advecting a strong contact discontinuity in the form of an overdense square that is in pressure equilibrium with the background. This tests the ability of our code to cope with a physical discontinuity that is not self-steeping, unlike a shock, i.e.~once the contact discontinuity is (excessively) broadened by numerical viscosity, it will invariably retain the acquired thickness. In fact, the advection errors inherently present in any Eulerian mesh-based numerical method will continue to slowly broaden a moving contact discontinuity with time, in contrast to Lagrangian methods, which can cope with this situation in principle free of any error.

A problem that starts with a perfectly sharp initial discontinuity furthermore tests the ability of our DG approach to cope with a situation where strong oscillatory behaviour is sourced in the higher order terms, an effect that is especially  strong  if the motion is supersonic and the system's state is characterized by large discontinuities. Here any naive implementation that does not include any type of limiter or artificial viscosity terms will invariably crash due to the occurrence of unphysical values for density and/or pressure. The square advection problem is thus also a sensitive stability test for our high-order Discontinuous Galerkin method.

In our test we follow the setup-up of \citet{Schaal2015}, but see also \citet{Hopkins2015} for a discussion of results obtained with particle-based Lagrangian codes. In 2D, we consider a domain $[0,1]^2$ with pressure $P=2.5$ and $\gamma=7/5$. The density inside the central square of side-length $0.5$ is set to $\rho=4$, and outside of it to $\rho=1$. A velocity of $v_x=100$ is added to all the gas, and in the $y$-direction we add $v_y=50$. We simulate until $t=1.0$, at which point the pulse has been advected 100 times through the periodic box in the $x$-direction and half that in the $y$-direction, and it should have returned exactly to where it started. We have also run the same test problem generalized to 3D, with an additional velocity of $v_z = -25$ in the $z$-direction, and as well in 1D, where only the motion in the $x$-direction is present. In general, the multi-dimensional tests behave very similarly to the one-dimensional tests, with the size of the overall error being determined by the largest velocity. For simplicity, we therefore here restrict ourselves in the following to report explicit results for the 1D case only.

In Figure~\ref{fig:square_advection_shape}, we show the density profile of the pulse at $t=1.0$ when 10 cells per dimension are used, for different DG expansion orders $p$. A second-order accurate method, $p=1$, which is equivalent or slightly better than common second-order accurate finite volume methods \citep[see also][]{Schaal2015} has already washed out the profile substantially at this time. Already order $p=2$ does substantially better, with $p=3$ results starting to resemble a top hat profile. The 10th order run with $p=9$ is able to retain the profile very sharply, albeit with a small amount of ringing right at the discontinuities. Similarly to our results for shocks, we thus find that our code is able to make good use of higher order terms if they are available in the expansion basis. Applying simple limiting schemes such as minmod in the high-order case is in contrast prone to lose much of the benefit of high-order when string discontinuities are present in the simulation, simply because these schemes tend to discard subcell information beyond linear slopes. We also show a $p=9$ order run with higher artificial viscosity injection in regions with wiggles by using a lower $S_{\rm onset} = 10^{-6}$. Spurious oscillations get dampened at the cost of a slightly wider shock front.

Comparing our results in the bottom part of Fig.~\ref{fig:square_advection_shape} to the DG implementation with posteriori correction by \citet[][their Figs.~5~and~7a]{vilar2019}, we can see that both methods successfully capture the sharp transition in a sub-cell fashion. For this comparison it is worth noting that in our case the shock is resolved within one cell, while in the setup of \citet{vilar2019} the discontinuity occurs at the border of two cells.

To look more quantitatively at the errors, we show in Figure~\ref{fig:square_advection_L1} the L1 error norm of the density field as a function of time, for all orders from $p=0$ to $p=9$. We see that the lowest order does very poorly on this problem, due to its large advection errors. In fact, $p=0$ loses the profile completely after about 10 box transitions, yielding a uniform average density throughout the whole box. When one uses higher order, both the absolute error at any given time  but also the rate of residual growth of the error with time drop progressively. The latter can be described by a power-law $L1 \propto t^n$, with a slope $n$ that we measure to be just 0.028 for $p=9$, while it is still 0.335 for a second-order, $p=1$ method. The longer a simulation runs, the larger the accuracy advantage of a high-order method over lower-order methods thus becomes.

\section{Kelvin-Helmholtz instabilities}
\label{SecKH}

Simulations of the Kelvin-Helmholtz~(KH) instability have become a particularly popular test of hydrodynamical codes \citep[e.g.][]{Price2008, Springel2010, Junk2010, Valcke2010, Cha2010, Berlok2019, Tricco2019, Borrow2022}, arguably initiated by an influential comparison of SPH and Eulerian codes by \citet{Agertz2007}, where significant discrepancies in the growth of the perturbations in different numerical methods had been identified. One principal complication, however, is that for initial conditions with an arbitrarily sharp discontinuity the non-linear outcome is fundamentally ill-posed at the discretized level \citep[e.g.][]{Robertson2010, McNally2012}, because for an ideal gas the shortest wavelengths have the fastest growth rates, so that one can easily end up with KH billows that are seeded by numerical noise at the resolution limit, rendering a comparison of the non-linear evolution between different methods unreliable. Furthermore, in the inviscid case, the non-linear outcome is fundamentally dependent on the numerical resolution so a converged solution does not even exist.

\begin{figure*}
\begin{center}
	\resizebox{18cm}{!}{\includegraphics{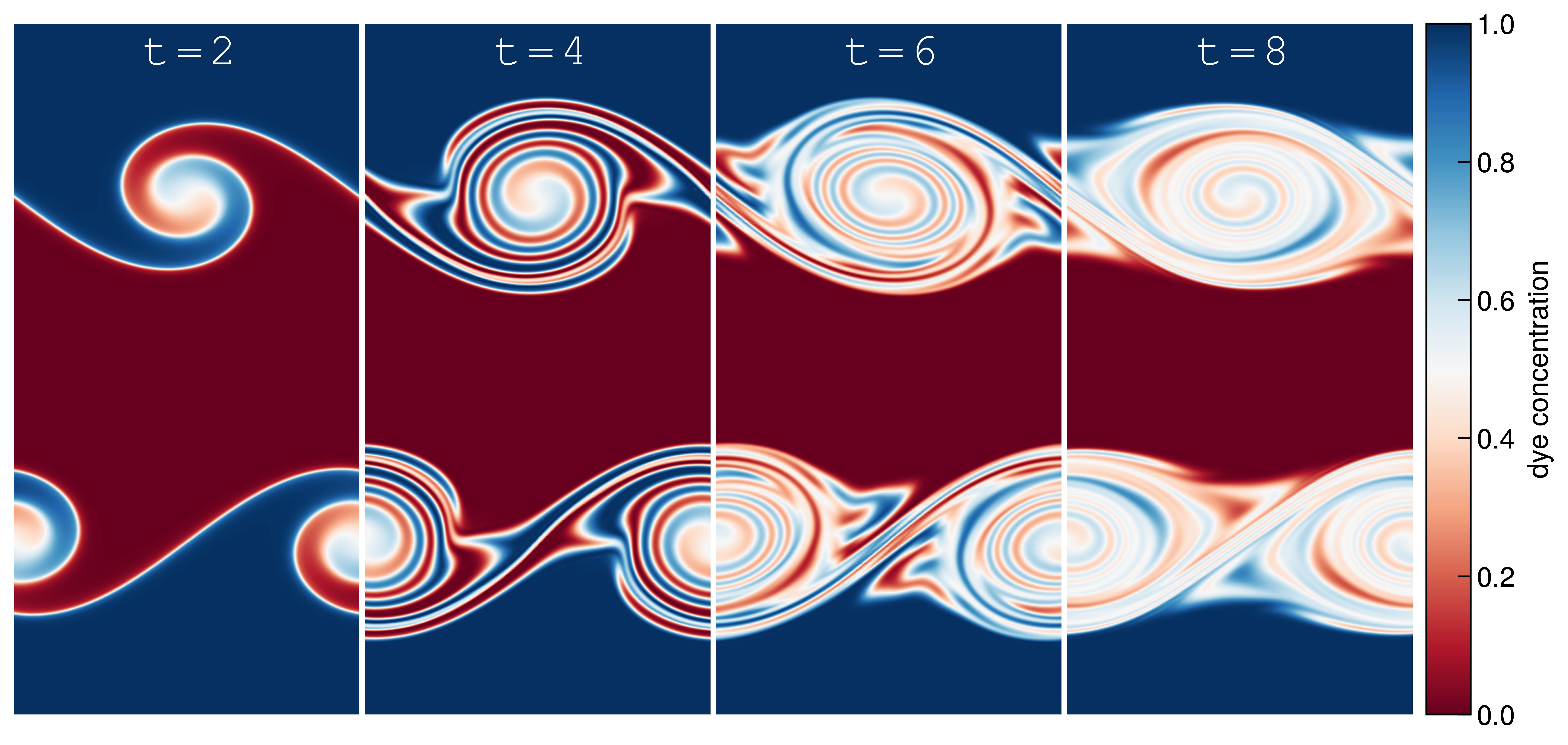}}
	\end{center}
    \caption{Time evolution of the dye concentration in a Kelvin-Helmholtz simulation using 64 DG-cells along the $x$-range $[0,1]$, at order $p=5$, using a viscosity setting of ${\rm Re}=10^5$ and $\Delta \rho / \rho_0 = 0$.}
    \label{fig:kh_different_times}
\end{figure*}

\begin{figure*}
\begin{center}
    \resizebox{14cm}{!}{\includegraphics{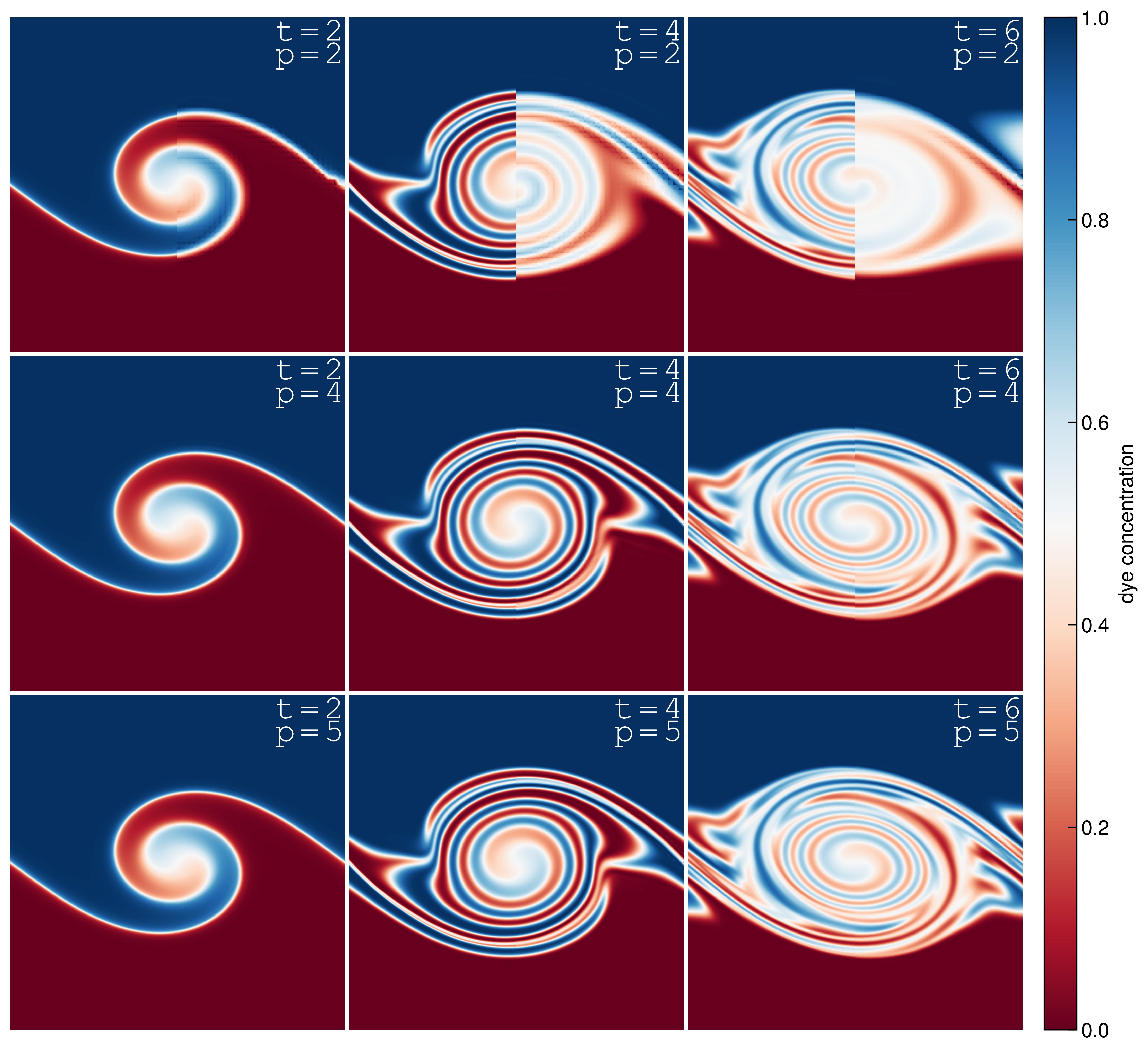}
}
\end{center}
    \caption{Dye concentration in Kelvin-Helmholtz simulations, using ${\rm Re}=10^5$ and $\Delta \rho / \rho_0 = 0$, compared at fixed grid resolution but different times $t$ and order $p$. Each of the nine panels shows the high-resolution {\small DEDALUS} reference result \citep{Lecoanet2016} in the left half, and our DG result (at different order $p$ as labelled) in the right half. 
    All DG simulations were done with $N_c=64$ grid cells.
    }
    \label{fig:kh_time_vs_order_lecoanet_comparison}
\end{figure*}

\citet{Lecoanet2016} have therefore argued that using smooth initial conditions across the whole domain combined with an explicit physical viscosity is a better choice, because this allows in principle converged numerical solutions to be reached. We follow their approach here, and in particular compare to the reference solution determined by \citet{Lecoanet2016} using the  spectral code {\small DEDALUS} \citep{Burns2020} at high resolution.

Specifically, following \citet{Lecoanet2016}  we adopt as initial conditions a shear flow with a smoothed density and velocity transition:
\begin{eqnarray}
\begin{aligned}
\rho(x,y) &=1+\frac{\Delta \rho}{\rho_{0}}  \frac{1}{2}\left[\tanh \left(\frac{y-y_{1}}{a}\right)-\tanh \left(\frac{y-y_{2}}{a}\right)\right], \\
v_{x}(x,y) &= u_{\text {flow }} \left[\tanh \left(\frac{y-y_{1}}{a}\right)-\tanh \left(\frac{y-y_{2}}{a}\right)-1\right], 
\end{aligned}
\end{eqnarray}
with $u_\textrm{flow} = 1$, $a = 0.05$, $y_1 = 0.5$ and $y_2 = 1.5$ in a periodic domain with side length $L=2$. This is perturbed with a small velocity component in the $y$-direction to seed a KH billow on large scales:
\begin{equation}
v_{y}(x,y) =A \sin (2 \pi x) \left[\exp \left(-\frac{\left(y-y_{1}\right)^{2}}{\sigma^{2}}\right)+\exp \left(-\frac{\left(y-y_{2}\right)^{2}}{\sigma^{2}}\right)\right] ,
\end{equation}
where $A = 0.01$ and $\sigma = 0.2$ is chosen. The pressure is initialized everywhere to a constant value, $P(x,y) =P_{0}$, with $P_0 = 10$. With these choices, the flow stays in the subsonic regime with a Mach number $\mathcal{M} \sim 0.25$. 

The free parameter $\Delta \rho / \rho_0$ describes the presence or absence of a density ``jump'' across the two fluid phases that stream past each other. By adding a passive tracer field
\begin{equation}
\begin{aligned}
c(x,y) &=\frac{1}{2}\left[\tanh \left(\frac{y-y_{2}}{a}\right)-\tanh \left(\frac{y-y_{1}}{a}\right)+2\right]
\end{aligned}
\end{equation}
to the initial conditions, we can study the KH instability also easily for the case of a vanishing density jump. In fact, we shall focus on the case $\Delta \rho / \rho_0=0$ here as it is free of the particularly subtle inner vortex instability in the late non-linear evolution of the KH problem \citep{Lecoanet2016}, which further complicates the comparison of different codes. 

To realize the above initial conditions we evaluate them within each cell of our chosen mesh at multiple quadrature points in order to project them onto our DG basis. We perform this initial projection using a Gauss integration that is 2 orders higher than that employed in the run itself. 
This ensures that integration errors from the projection of the initial conditions onto our DG basis are subdominant compared to the errors incurred by the time evolution, and are thus unimportant.

We choose identical values for shear viscosity $\nu$, thermal diffusivity $\chi$,  and dye diffusivity $\eta$. Below, we mostly focus on discussing results for a Reynolds number ${\rm Re} = 10^5$ for which  we set $\nu = \chi = \eta = 2 u_{\rm flow} / {\rm Re} = 2 \times 10^{-5}$. We have also carried out simulations with a higher Reynolds number ${\rm Re} =  10^6$, obtaining qualitatively similar results, although these simulations require higher resolution for convergence and thus tend to be more expensive.

\subsection{Visual comparison}

A visual illustration of the time evolution of the dye concentration for a simulation with ${\rm Re} = 10^5$ and $\Delta \rho / \rho_0 = 0$ is shown in Fig.~\ref{fig:kh_different_times}. In this calculation, 64 DG cells were used to cover the $x$-range $[0,1]$, which is the relevant number to compare to the resolution information in \citet{Lecoanet2016}. Expansion order $p=6$ has been used in this particular run. It is nicely seen that the KH billow seeded in the initial conditions is amplified in linear evolution until a time $t \sim 1-2$, then it rolls up multiple times in a highly non-linear evolution, before finally strong mixing sets in that progressively washes out the dye concentration throughout the vortex.

Upon visual inspection, this time evolution compares very closely to that reported by \citet{Lecoanet2016}. In Figure~\ref{fig:kh_time_vs_order_lecoanet_comparison} we make this comparison more explicit by showing results obtained for different order $p$ at a number of times `face-to-face' with their reference simulation. In each of the panels, the left half of the picture contains the {\small DEDALUS} result at resolution $3096\times 6144$, while the right half gives our results at $64\times 128$ resolution, but with different orders $p$. We have deliberately chosen this modest resolution for this comparison in order to allow some differences to be seen after all. They show up clearly only at second-order in the top row, while at $p=4$ they are only discernible at times $t=4$ and $t=6$ as faint discontinuities at the middle of the images, where the result from  {\small DEDALUS} meets that from our DG code. Already by $p=5$, visual inspection is insufficient to identify clear differences. We note that for higher DG grid resolutions, this becomes rapidly extremely difficult already for lower orders.

\begin{figure}
	\includegraphics[width=\columnwidth]{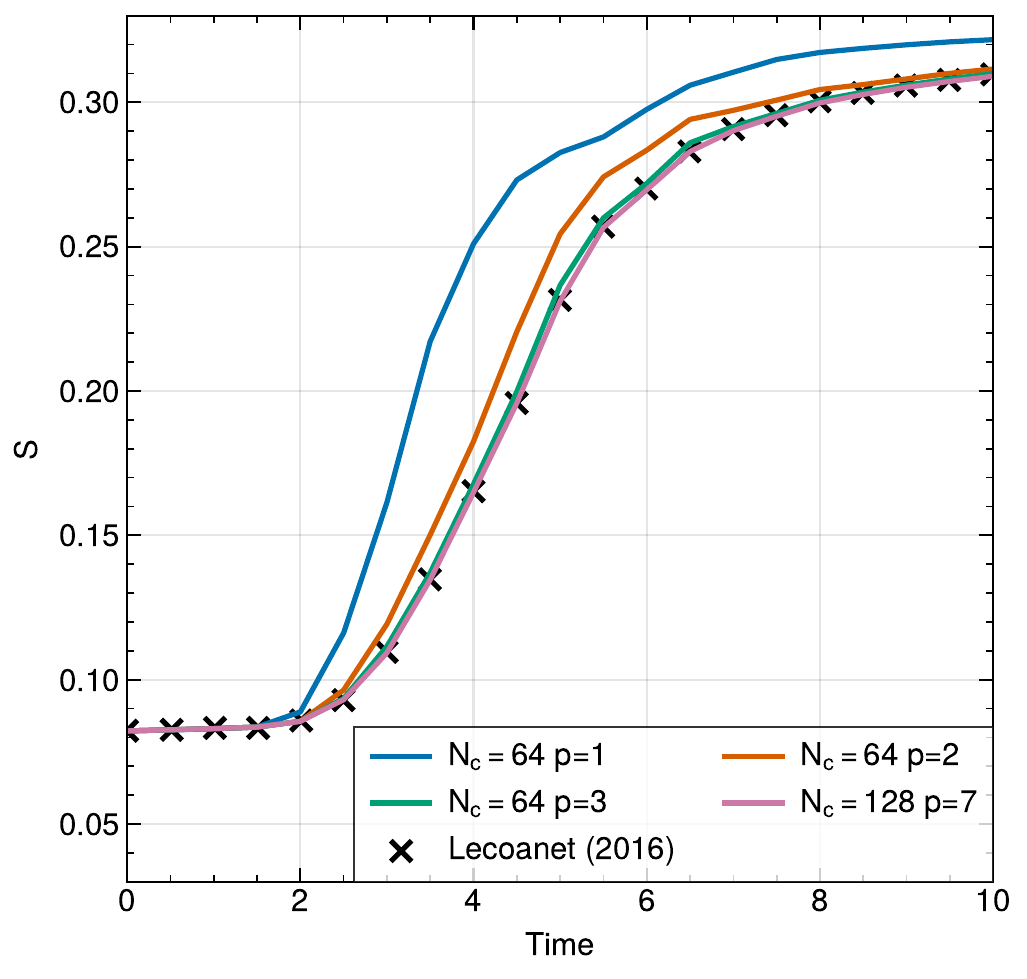}
    \caption{Volume integrated dye entropy as a function of time. We show our DG simulation results with 64 cells using orders $p=1$ to $p=3$, and a calculation with 128 cells and order $p=7$. All simulations were ran with ${\rm Re}=10^5$ and a density jump $\Delta \rho / \rho_0 = 0$. Already the run with 64 cells and $p=3$ shows an essentially converged result; still better resolutions yield perfect agreement with the very high resolution results obtained by \citet{Lecoanet2016} with the  {\small DEDALUS} and  {\small ATHENA} codes.
    }
    \label{fig:khsmooth_jump_re5}
\end{figure}

\subsection{Dye entropy}
\label{SecKHdye}

An interesting more quantitative comparison of our simulations to those of \citet{Lecoanet2016} can be carried out by considering the evolution of the passive scalar  ``dye'' in some detail. The technical implementation of this passive tracer is described in Section~\ref{SecPassiveScalar}.

A dye entropy per unit mass can be defined as $s = -c \ln c$, and its volume integral is the dye entropy
\begin{equation}
    S = \int \rho s \,{\rm d}V,
\end{equation}
which can only monotonically increase with time. The dye entropy can be viewed as a useful quantitative measure for the degree of mixing that occurs as a result of the non-linear KH instability. To guarantee an accurate measurement of the dye entropy we perform the integral above at two times higher spatial order than employed in the actual simulation run. We also note than when computing the dye entropy we use our entire simulation domain (although left and right halves give identical values), and we then normalize to half of the volume to make our values directly comparable to those of \citet{Lecoanet2016}.

In Figure~\ref{fig:khsmooth_jump_re5}, we show measurements of the dye entropy evolution for several of our runs, compared to the converged results obtained by \citet{Lecoanet2016} consistently with the {\small DEDALUS} and {\small ATHENA} codes. We obtain excellent agreement already for 64 cells and order $p=4$, corresponding to 256 degrees of freedom per dimension. Our under-resolved simulations with fewer cells and/or degrees of freedom show an excess of mixing and higher dye entropy, as expected.

We note that \citet{Tricco2019} have also studied this same reference problem using SPH. Interestingly, they find that simulations that are carried out at lower resolution than required for (approximate) convergence show an underestimate of dye mixing, marking an important qualitative difference to the mesh-based computations. The SPH simulations also require a substantially higher number of resolution elements to obtain an approximately converged result. \citet{Tricco2019} get close to achieving this for the dye concentration by using 2048 particles per dimension, but even then the dye entropy of their result falls slightly below the converged result at $t=8$.

\subsection{Error norm}

\begin{figure}
	\includegraphics[width=\columnwidth]{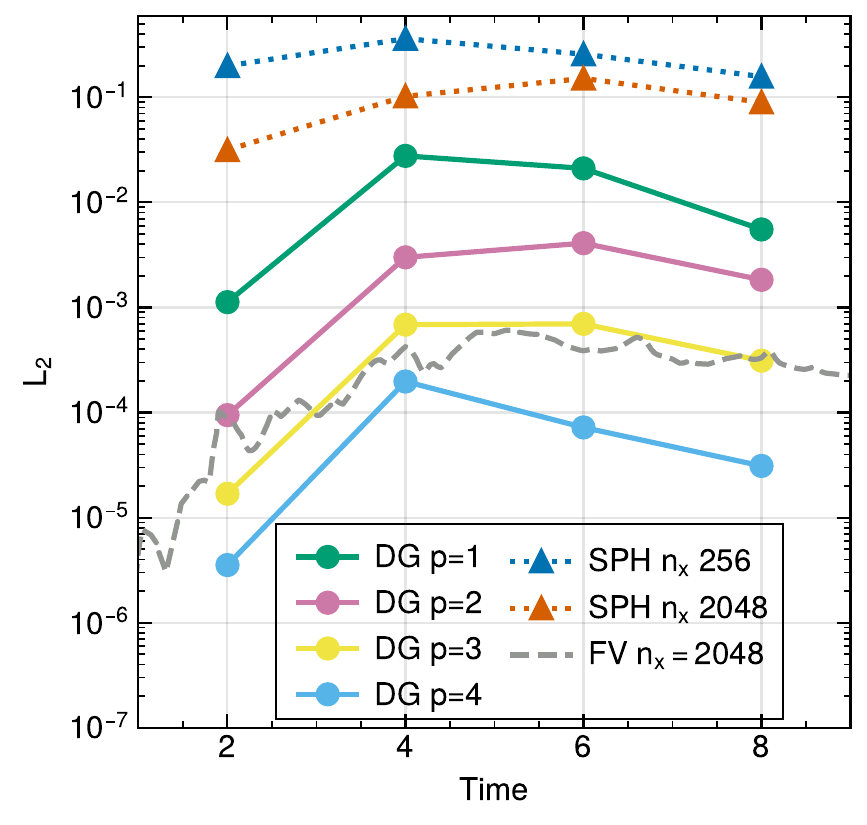}
    \caption{Volume-averaged $L_2$-error norm of the difference in the dye concentration relative to a high-resolution spectral result as a function of time, for a set of DG simulations carried out with 64 cells and different expansion order $p=1$ to $p=4$ (as labelled), for ${\rm Re}=10^5$ and a density jump $\Delta \rho / \rho_0 = 0$. The DG results are presented with filled circles at the four available output times of the spectral simulation, the connecting lines are there simply to guide the eye. Similarly, we include SPH results by \citet{Tricco2019} as triangles at two different resolutions. Finally, the dashed line refers to the result obtained by \citet{Lecoanet2016} using the finite-volume code {\small ATHENA} with 2048 cells. }
    \label{fig:kh_l2_over_time}
\end{figure}

\begin{figure}
	\includegraphics[width=\columnwidth]{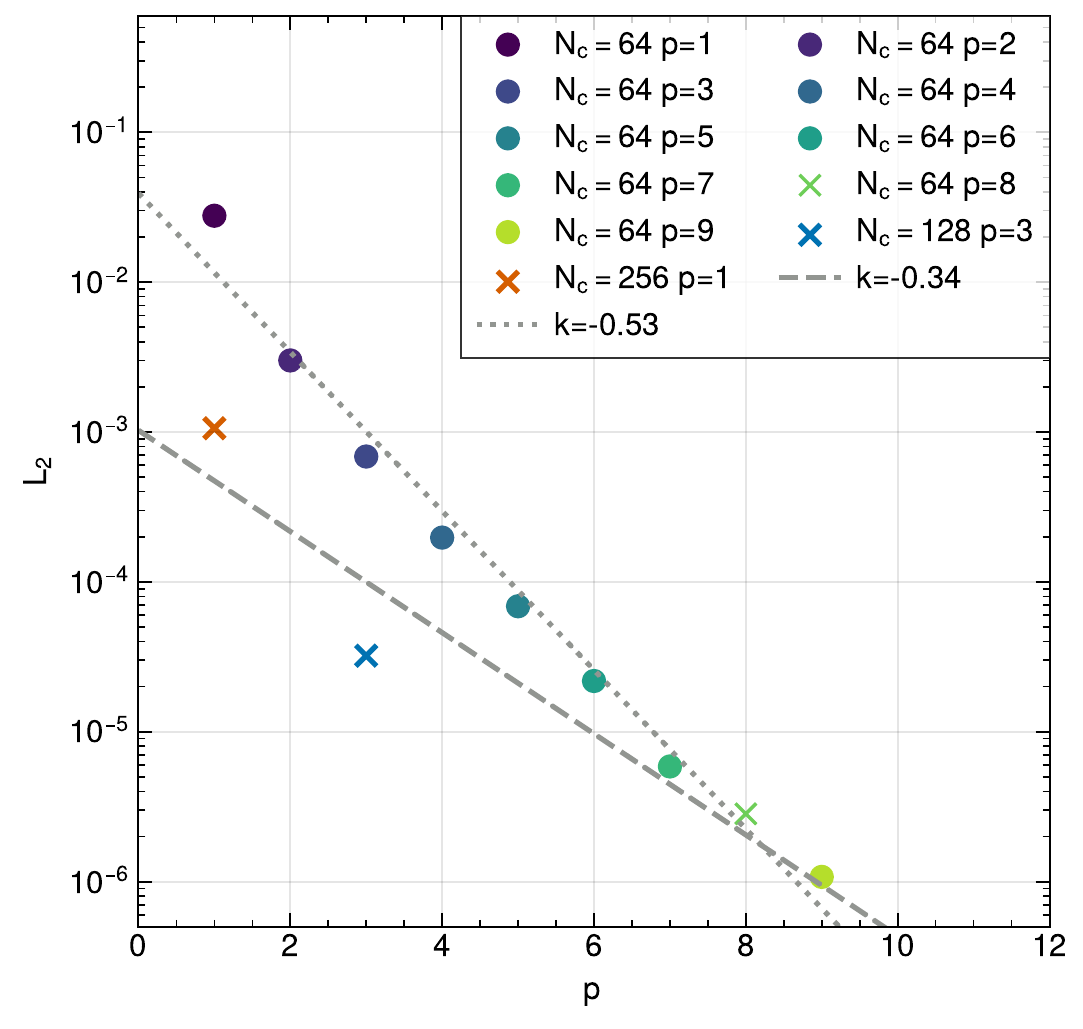}
    \caption{Volume-averaged $L_2$ error norm of the dye concentration field  as a function of DG order $p$ for a set of simulations with ${\rm Re}=10^5$ and a density jump $\Delta \rho / \rho_0 = 0$ at $t=4$. The circles show simulations with $N_c=64$ cells with progressively increasing order $p$ (the run with $p=8$ is shown with a cross symbol while still being a member of the sequence of simulations with circles). The crosses highlight three simulations with the same number of degrees of freedom, reached with different combinations of $N_c$ and $p$. The dotted line is a fit showing the rapid convergence we achieve with increasing order $p$ at $N_\textrm{c}=64$. The dashed line indicates the convergence rate for three simulations with equal number of degrees of freedom, as we increase the order. Among the three runs with an equal number of degrees of freedom, the one with the highest order $p$ achieves the  lowest $L_2$-norm.
    }
    \label{fig:kh_l2_at_t4}
\end{figure}

Finally, we consider a direct comparison of the dye entropy fields obtained in our simulations to the {\small DEDALUS} reference solution made publicly\footnote{https://doi.org/10.5281/zenodo.5803759} available by \citet{Lecoanet2016} at a grid resolution of $3096 \times 6144$ points. 
To perform a quantitative comparison, we consider the $L_2$-norm of the
difference in the dye fields, defined as
\begin{equation}
L_2 = \left[\frac{1}{V} \int  \left(c_{\rm DG} - c_{\rm Lecoanet}\right)^2 {\rm d}V \right]^{1/2} .
\end{equation}

In Figure~\ref{fig:kh_l2_over_time} we show first the time evolution of the $L_2$-norm, for DG simulations carried out with 64 cells and orders $p=2$ to $p=5$. We also include results reported  by  \citet{Lecoanet2016} for the {\small ATHENA} mesh code at a resolution of 1024 cells, as well as SPH results by \citet{Tricco2019} at particle resolutions of 256 and 2048, respectively. Our $p=4$ results with 64 cells are already as good as  {\small ATHENA} with 2048 cells, demonstrating that far fewer degrees of freedom are sufficient when a high order method is used for this smooth problem. In contrast, a relatively noisy method such as SPH really struggles to obtain truly accurate results. Even at the 2048 resolution, the errors are orders of magnitude larger than for the mesh-based methods, and the sluggish convergence rate of SPH will make it incredibly costly, if possible at all, to push the error down to the level of what our DG code, or {\small ATHENA}, comparably easily achieve.

In Figure~\ref{fig:kh_l2_at_t4}, we examine the error as a function of the employed DG expansion order. For a fixed number $N_c=64$ of cells, we show the $L_2$ error at time $t=4$, for orders $p=1$ up to $p=9$. It is reassuring that we again find exponential convergence for this problem, where the error drops approximately linearly with $p$ on this log-linear plot. This demonstrates that we can fully retain the ability to converge at high-order for our compressible Navier-Stokes solver, which is additionally augmented with thermal and dye diffusion processes. We consider this to be a very important validation of our numerical methodology and actual code implementation. 

Another interesting comparison is to consider simulations that have an equal number of degrees of freedom, but different cell numbers and expansion orders. In the figure (marked with crosses), we also include results for the three cases $N_c=64/p=8$, $N_c=128/p=4$, and $N_c=256/p=2$, which all have the same number of degrees of freedom per dimension. Strictly speaking, the higher order ones have actually slightly less, given that the number $N^{\rm 2D}(p) = p(p+1)/2$ of degrees of freedom per cell is slightly less than $p^2$ for $p>1$, see Equation~(\ref{eqn:N2D}). Regardless, the run with $N_c=64$ clearly has the lowest error. This confirms once more that for a smooth problem it is typically worthwhile in terms of yielding the biggest gain in accuracy to invest additional degrees of freedom into higher order rather than additional cells.

\begin{figure}
	\includegraphics[width=\columnwidth]{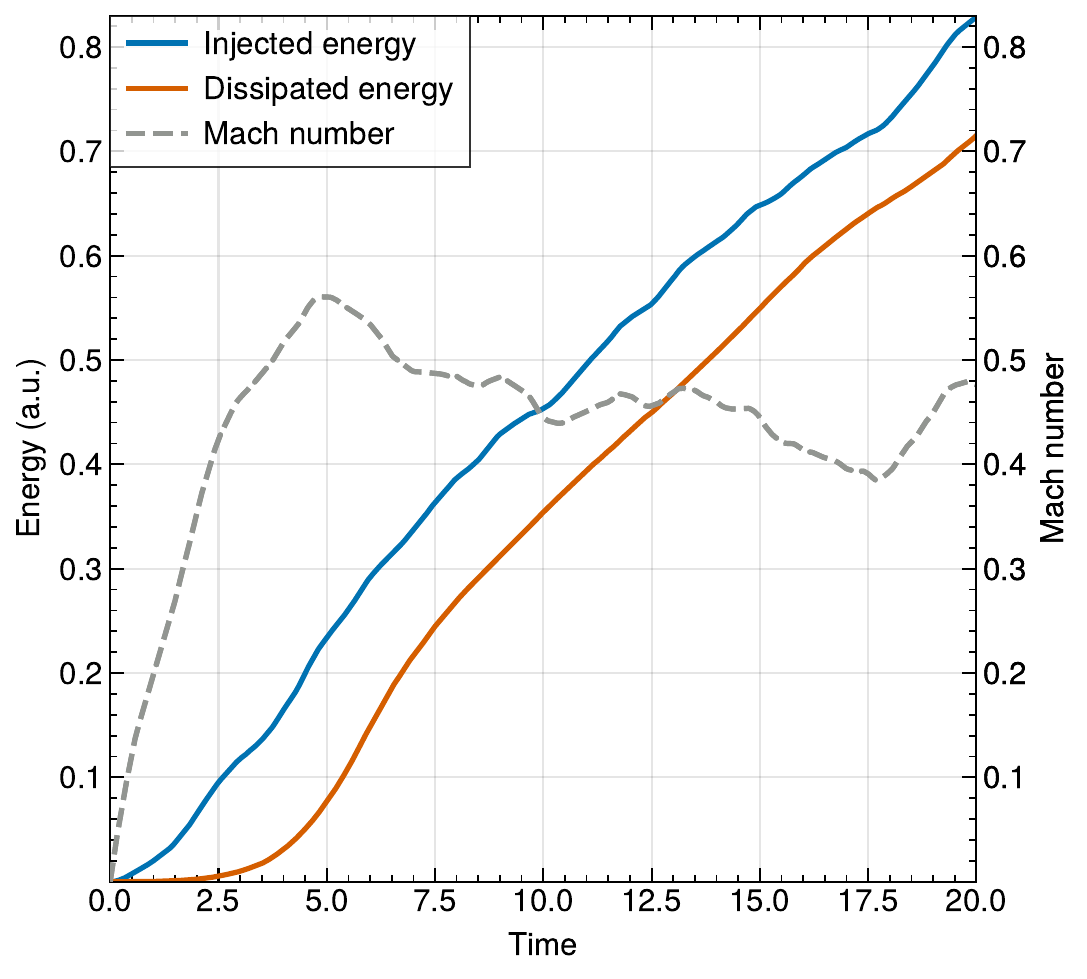}
    \caption{Cumulative injected and dissipated energy, as well as  global Mach number, as a function of time in one of our driven turbulence simulations. The gas is initially at rest, and put into motion by the driving. Eventually, energy injection is balanced by dissipation in a time-averaged fashion, and the difference between the cumulative injected and dissipated energy is reflected in the kinetic energy as measured by the Mach number.}
    \label{fig:turbulence_energy_and_mach}
\end{figure}

\begin{figure}
	\includegraphics[width=\columnwidth]{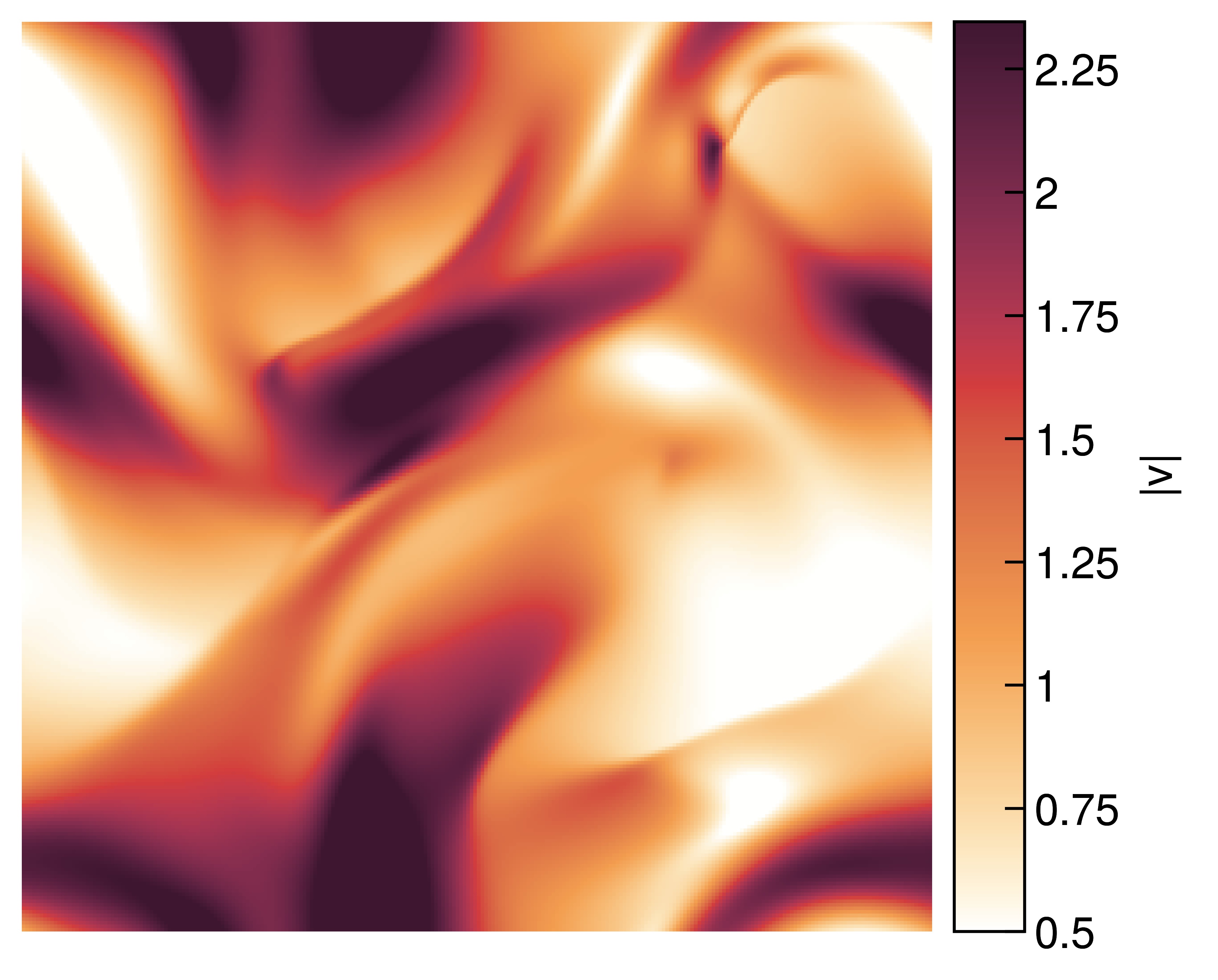}
    \caption{Two-dimensional slice through a driven, isothermal, subsonic 3D turbulence simulation depicting the  velocity amplitude \mbox{$|v| = ({v_x^2 + v_y^2 + v_z^2})^{1/2}$} at $t=20.48$, for a simulation with $N_c=128$, $p=4$, and ${\rm Re} = 10^5$.}
    \label{fig:turbulence_velocity_amplitude}
\end{figure}

\begin{figure*}
	\resizebox{16cm}{!}{\includegraphics{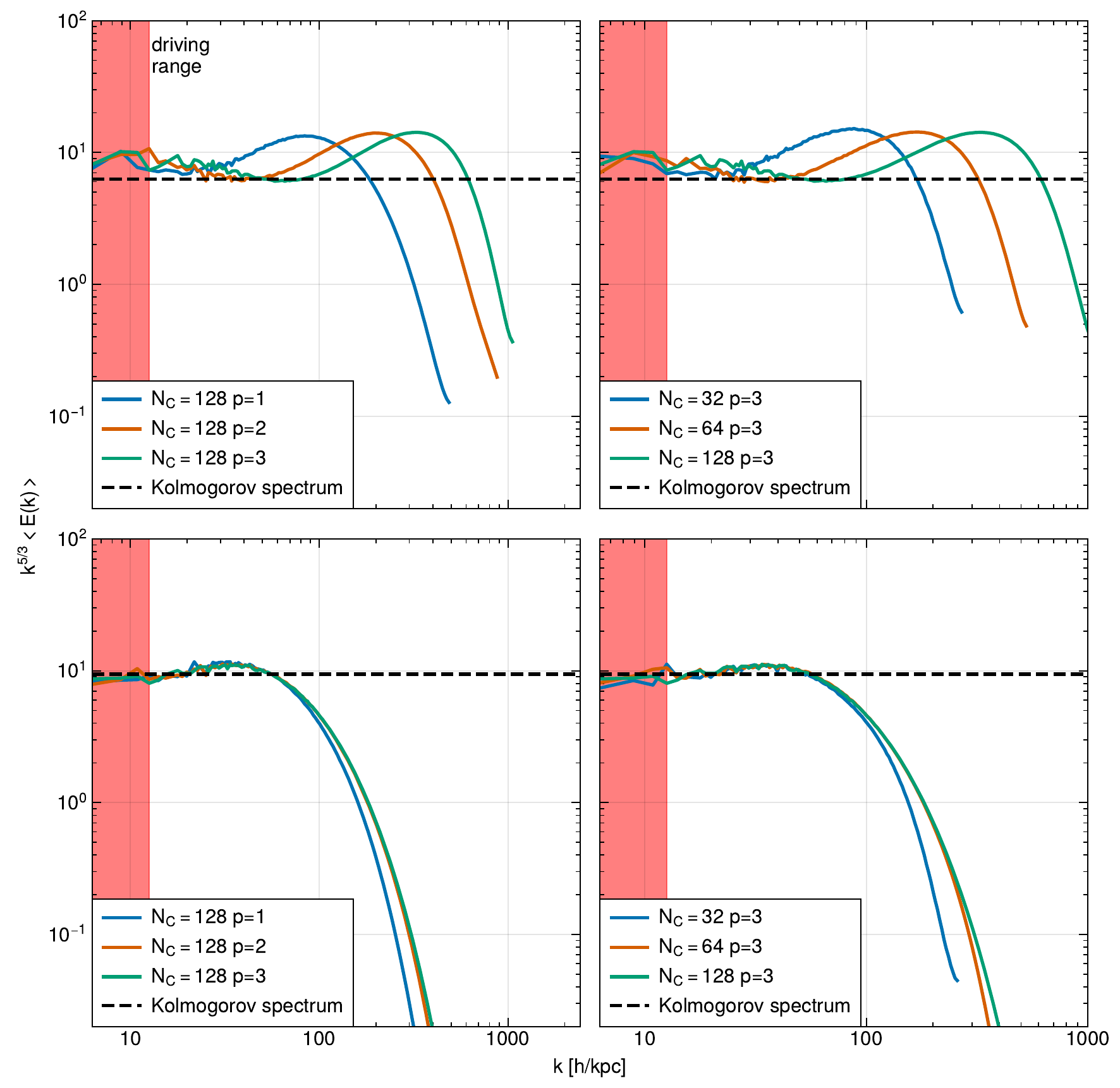}}
    \caption{Compensated velocity power spectra of driven turbulence simulations as a function of wavenumber for varying numbers of cells, and varying spatial order. The panels in the top row show simulations where the Euler equations were solved, whereas the bottom two panels give results where the full compressible Navier-Stokes equations with a prescribed physical viscosity were used. The region marked with a red shade is the driving range.}
    \label{fig:turbulence_euler_ns_order_vs_nc}
\end{figure*}

\section{Driven sub-sonic turbulence}
\label{SecTurbulence}

The phenomenon of turbulence describes the notion of an unsteady, random flow that is characterized by the overlap of swirling motions on a variety of scales \citep[e.g.][]{Pope2000}. In three-dimensions, one finds that if fluid motion is excited on a certain scale (the injection scale) it tends to decay into complex flow features on ever smaller scales, helped by fluid instabilities such as the Kelvin-Helmholtz instability. Eventually, the vortical motions become so small that they are eliminated by viscosity on the so-called dissipation scale. If the injection of kinetic energy on large scales persists and is quasi-stationary, a fully turbulent state develops which effectively exhibits a transport of energy from the injection to the dissipation scale. For incompressible isotropic, subsonic turbulence, the statistics of velocity fluctuations in such a turbulent flow are described by the Kolmogorov velocity power spectrum, which has a power law shape in the inertial range, and a universal shape in the dissipative regime.

For astrophysics, turbulence plays a critical role in many environments, including the intracluster medium, the interstellar medium, or the buoyantly unstable regions in stars. Numerical simulations need to be able to accurately follow turbulent flows, for example in order to correctly describe the mixing of different fluid components, or the amplification of magnetic fields. However, this is often a significant challenge as the scale separation between injection and dissipation scales in  astrophysical settings can be extremely large, while for three-dimensional simulation codes it is already difficult to resolve even a moderate difference between injection and dissipation scales. In addition, most astrophysical simulations to date rely on numerical viscosity exclusively instead of including an explicit physical viscosity, something that can in principal modify the shape of the dissipative part of the turbulent power spectrum, thereby creating turbulent velocity statistics that differ from the expected universal form because they are directly affected by aspects of the numerical method. 

Of course, the general accuracy of a numerical method is also important for how well turbulence can be represented. For example, \citet{BauerSpringel2012} have pointed out that the comparatively large noise in SPH makes it  difficult for this technique to accurately represent subsonic turbulence. While this can in principle be overcome with sufficiently high numerical effort, it is clear that methods that have a low degree of numerical viscosity combined with the ability to accurately account for physical viscosity should be ideal for turbulence simulations. Our DG approach has these features, and especially in the regime of subsonic turbulence, where shocks are expected to play only a negligible role, the  DG method should be particularly powerful.

This motivates us to test this idea in this section by considering isothermal, subsonic, driven turbulence in periodic boxes of unit density. The subsonic state refers to the average kinetic energy of the flow in units of the soundspeed, as measured through the Mach number. Instead of directly imposing an isothermal equation of state, we simulate gas with a normal ideal gas equation of state and reset the temperature every timestep such that a prescribed sound speed is retained. We have checked that this does not make a difference for any of our results, but this approach allows us to use our approximate, fast HLLC Riemann solver instead of having to employ our exact, but slower isothermal Riemann solver.

\subsection{Driving}

To create the turbulence, we drive fluid motions on large scales. To do this consistently at high order, we add a source function $\boldsymbol{s}(\boldsymbol{x}, t)$ to the right-hand side of the Euler equations, both in the momentum equation and as work function $\boldsymbol{s} \cdot \boldsymbol{v}$ in the energy equation. These source terms have to be integrated with Gaussian quadrature over cell volumes to retain the high-order discretization.

For setting up the driving field $\boldsymbol{s}(\boldsymbol{x}, t)$, we follow standard techniques as implemented in \citet{BauerSpringel2012}, which in turn are directly based on \citet{Schmidt2006, Federrath2008, Federrath2009}.  The acceleration field is constructed in Fourier space by populating modes in the narrow range $2\pi/L \le k \le 2\times 2\pi/L$, with amplitudes scaled to follow $\propto k^{-5/3}$ over this range. The phases of the forcing modes are drawn from an Ornstein–Uhlenbeck process. They are periodically updated whenever a time interval $\Delta t$ has elapsed, while keeping a temporal correlation over a timescale $t_s$ with the previous phases. This effectively yields a smoothly varying, random driving field. Our specific settings for update frequency, coherence timescale and distribution function for drawing the driving phases are as in \citet[][their table 1, left column]{BauerSpringel2012}.

We here also restrict ourselves to include only solenoidal driving, i.e.~we project out all compressive modes in Fourier space by a Helmholtz decomposition. Specifically, if $\boldsymbol{s}$ is the principal acceleration field constructed in the above fashion, we project it as
\begin{equation}
\boldsymbol{\hat s}(\boldsymbol{k}) = \left( \delta_{ij} - \frac{k_i k_j}{\boldsymbol{k}^2}\right)  \boldsymbol{s}(\boldsymbol{k})
\end{equation}
in Fourier space to end up with an acceleration field $\boldsymbol{\hat s}$ that is free of compressive modes, which would only produce a spectrum of additional sound waves in our subsonic case. 

\subsection{Results for subsonic turbulence}

All our turbulence simulations are started with gas of uniform density at rest. We monitor the average kinetic energy, as well as the total cumulative injected kinetic energy and the total cumulative dissipated energy, allowing us to verify the establishment of a quasi-stationary state. An example for this is shown in Figure~\ref{fig:turbulence_energy_and_mach}, where we illustrate the build-up of the turbulent state in terms of the total energies. There is an initial ramp up phase of the turbulence until $t\sim 5$, during which the Mach number grows nearly linearly to its final quasi-stationary time-averaged value of ${\cal M} \simeq 0.47.$ The cumulative injected energy grows approximately linearly with time, whereas the dissipated energy tracks it with a time lag, because the initial evolution until $t\sim 2.5$ does not yet show any significant dissipation. The difference between the injected and dissipated energies is the current kinetic energy of the gas, and thus is effectively given by the Mach number.

In Figure~\ref{fig:turbulence_velocity_amplitude}, we show a visual example of the quasi-stationary turbulent state established after some time, here simulated with $N_c=128$ cells and order $p=4$. The slice through the magnitude of the velocity field  illustrates the chaotic structures characteristic of turbulence. Even though there are some steep gradients in the velocity field, the velocity varies smoothly overall,  reflecting the absence of strong shock waves in this subsonic case.

To statistically analyse the turbulent state we turn to measuring power spectra of the velocity field at multiple output times, and then consider a time-average spectrum to reduce the influence of intermittency. To calculate the final power spectrum of a simulation, we average over 64 velocity power spectrum measurements over the time interval $5.12<t<20.48$.

\subsubsection{Inviscid treatment of gas}

The behaviour of inviscid gas is described by the Euler equations of eqn.~(\ref{eq:euler}). Because of the simplicity of this model and the desire to run simulations with as little viscosity as possible to maximize the intertial range of the turbulence, it is a popular choice for the study of turbulence. For example, the largest driven turbulence simulation to date by \citet{Federrath2016LargestSimulation} were performed using inviscid gas, as well as many other studies in the field \citep[e.g.][]{2004astro.ph..7616S, BauerSpringel2012, 2016arXiv160209079B, Federrath2008, 2010A&A...512A..81F, 2010MNRAS.406.1659P}.

In the top two panels of Fig.~\ref{fig:turbulence_euler_ns_order_vs_nc}, we show  such simulations carried out with our DG code. In all such simulation, the energy injected at large scales follows the Kolmogorov spectrum and cascades from large to small scales. This part of the spectra is called the inertial range and it follows the $k^{-5/3}$ Kolmogorov spectrum closely, even though our gas is compressible and the density fluctuations for our Mach number are not negligible any more. Note that all our plots are compensated with a $k^{5/3}$ factor, such that  the Kolmogorov spectrum corresponds to a horizontal line. The extent of the inertial range is primarily determined by the total number of degrees of freedom in an inviscid simulation. However, as we transition from the inertial range to the dissipation portion of the spectra, a noticeable bump can be seen in which the spectrum significantly exceeds the power-law extrapolation from larger scales. As energy is being transferred from larger to smaller scales, creating ever smaller eddies, it eventually reaches scales at which the code cannot resolve smaller eddies any more. This leads to a build-up of an energy excess at this characteristic scale, until the implicit numerical viscosity terms become strong enough to dissipate away the arriving energy flux. This effect is commonly known in numerical studies of turbulence and referred to as the ``bottleneck'' effect. It should be pointed out that experimental determinations of turbulent velocity spectra also show a weak form of this effect \citep[see][and references therein]{Verma_2007}. \citet{kuchler_experimental_2019} later even measured the relation between the amplitude of the bump and $R_\lambda$ of the flow. The problem of numerical simulations of inviscid gas is however that the shape of the bump is determined by numerical details of the hydrodynamic code and that it is usually excessively pronounced.

The bottleneck effect cannot be fixed by using higher resolution, or higher order for that matter. Indeed, in the top two panels of Figure~\ref{fig:turbulence_euler_ns_order_vs_nc} we can see that the bottleneck moves to ever smaller scales with increasing cell number at a fixed spatial order, and similarly it moves towards smaller scales if we increase the spatial order of our method at a fixed number of cells. While both avenues of adding further degrees of freedom successfully widen the inertial range and push the dissipative regime to smaller scales, they unfortunately cannot  eliminate  the ``bump'' in the bottleneck, or address the equally incorrect detailed shape of the dissipation regime itself.  This detailed shape changes slightly as we vary the order $p$ because the precise way of how numerical dissipation interacts with the flow is modified by this, while in contrast increasing the number of cells leaves the shape unchanged, because this just moves the dissipation regime to smaller scales in a scale-invariant fashion. 

The only way around this and to get closer to velocity spectra seen in experimental studies of turbulence is to solve the full compressible Navier-Stokes equation, where the dissipative regime is set not by numerics, but by the physical viscosity of the gas itself. If this viscosity is large enough, it will effectively dissipate energy at scales larger than our numerical viscosity. We consider this case in the following subsection.

\subsubsection{Viscous treatment of gas}

We now consider driven turbulence results akin to the simulations just discussed, with the only difference being that we are now solving the full compressible Navier-Stokes equations as described in Sec.~\ref{SecNavierStokesEquations}. In the bottom two panels of Fig.~\ref{fig:turbulence_euler_ns_order_vs_nc}, we display compensated velocity power spectra with physical viscosity added. Such full Navier-Stokes simulations exhibit the proper behaviour of the ``bottlenect'' effect, as the location and shape of the bump become resolution-independent and do not depend on numerical code details any more. Such simulations are in the literature referred to as ``direct numerical simulations'' or DNS. Our code can achieve DNS for turbulence by either increasing the resolution or the spatial order, as is evident in the bottom two panels of Fig.~\ref{fig:turbulence_euler_ns_order_vs_nc}.

\begin{figure}
	\includegraphics[width=\columnwidth]{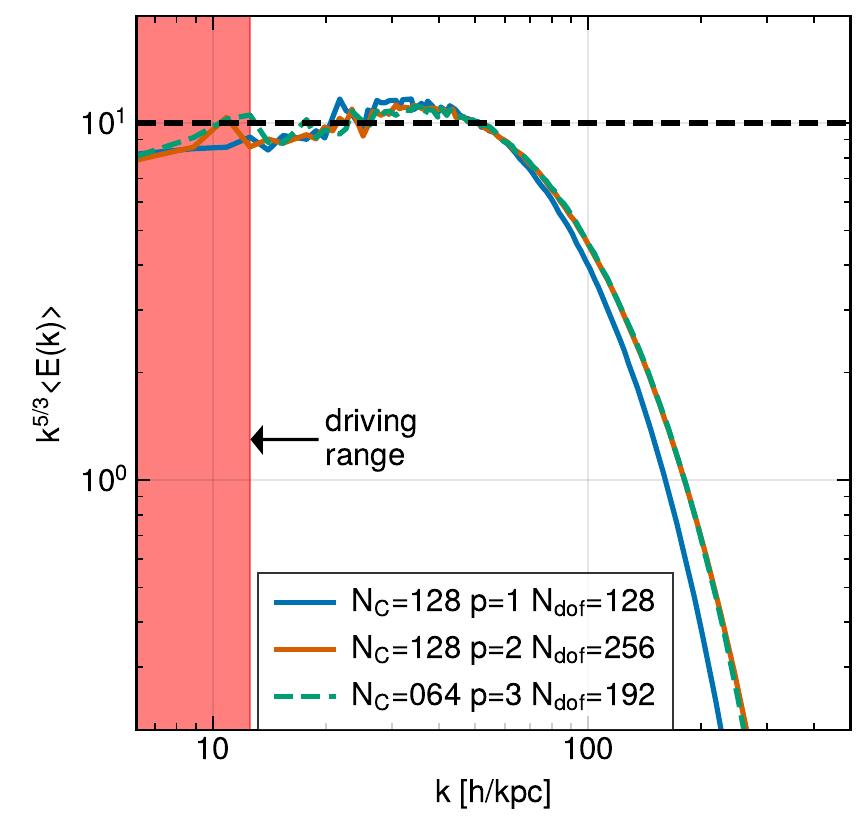}
    \caption{Compensated velocity power spectra as a function of wavenumber for a similar number of degrees of freedom, but varying the spatial order and the number of cells. The total wall-clock time for the simulation runs $128^3|p=2$, $128^2|p=3$, and $64^3|p=4$ on  16 A100 GPUs were 0.9, 3.9, and 1.8  hours, respectively. We note that one can keep the converged result obtained with $N_c=128$ and $p=3$ by going to fewer cells and higher order (the $N_c=64$ and $p=4$ run),  while still achieving a speed-up. 
}
    \label{fig:turbulence_order_over_cells}
\end{figure}

To determine if increasing the order of our method or its resolution is more beneficial, we compare three simulations with approximately the same number of degrees of freedom, but different resolutions and orders in Fig.~\ref{fig:turbulence_order_over_cells}. The orange line shows a run we can consider a converged DNS result with $N_\textrm{c}=128$ and $p=3$. A simulation with identical $N_\textrm{c}$ but lower $p$ in blue fails to fully converge. On the other hand, the green dashed line shows a simulation with \textit{eight times fewer} total number of cells, but at a higher spatial order. It has as many degrees of freedom as the simulation shown in blue, and yet its power spectra matches that of the simulation shown in orange. We can therefore conclude that running driven turbulence at higher order is preferable to increasing the cell resolution. Or to put it another way, if there is a limited number of degrees of freedom that can be represented due to memory constraints, it is better to ``spend'' the memory on higher $p$ than $N_\textrm{c}$. In the present case, a comparison of the wall-clock time between the high cell resolution and high order runs shows an about 2x faster calculation time at high order vs using a higher cell resolution. For even high order, this CPU-time advantage may not persist, but the memory advantage will. Given that turbulence simulations tend to be memory-bound, this in itself can already be a significant advantage.

\section{Code details}
\label{SecCode}

\subsection{Parallelization strategy}
\label{SecParallelizationStrategy}

Modern supercomputers consist by now of thousands to millions of computing cores, a trend which is bound to continue. Recently, however, the most significant gains in computational performance (measured in floating point operations per second -- FLOPS) have came from dedicated accelerator cards. These are most commonly, but not always, graphics processing units (GPUs) that have been repurposed to do general computational work. Accelerators achieve a large number of FLOPS by foregoing large, per-core caches and advanced control circuitry for single compute units, while at the same time they are able to execute large sets of threads concurrently in a data-parallel fashion.

Current GPU-accelerated computers typically consist of normal, CPU-equipped compute nodes that are outfitted with attached GPU cards. Utilising the power of both, CPUs and GPUs, efficiently with such heterogeneous machines is challenging. It requires not only a suitable subdivision of the work, but often also an algorithmic restructuring of the computations such that they can be mapped efficiently onto the massively parallel execution model of GPUs, as well as prescriptions for data placement and movement between the separate memory of CPU and GPUs. The problem becomes even harder when multiple compute nodes with distributed memory, each with their own GPUs, are supposed to work together on a tightly coupled problem. Efficient and scalable massively parallel codes for such machines must decompose the problem into multiple parts, distribute the parts among the available compute units, and only exchange data between various parts when really needed.

In the present version of our code {\small TENETGPU}\footnote{While our code is written from scratch for GPUs, its first version has been heavily inspired by the code {\small TENET} of \citet{Schaal2015}, hence we named ours {\small TENETGPU}. Source code: \url{https://bitbucket.org/Migelo/gpu_testbed}}, we address this by an implementation that can execute a given hydrodynamical problem flexibly either on one or several GPUs, on one or multiple CPU cores, or a mixture thereof. Independent on how GPUs and CPU-cores are distributed onto different compute nodes,  {\small TENETGPU} can in this way make use of whatever is available, up to extremely powerful systems such as the first exascale supercomputers that are presently put into service (which are GPU-accelerated, such as `Frontier', ranked the fastest in the world according to the Top500 list released May 30, 2022).

To achieve this flexibility, we split the mesh along the $x$-axis into different slabs, which can have different thickness, if desired. Each slab is either computed by a different GPU, or by one CPU core. The communication between slabs, which is realized with the Message Passing Interface (MPI), thus needs to happen along the $x$-dimension between neighboring slabs only, as all the needed data along the other two axes is locally available for the corresponding slab. The data that is communicated consists of surface states or surface fluxes at Gauss points needed for integrations over cell areas. For driving the GPU computations, each GPU requires a separate CPU core as well. For example, if one has a compute node with 32 cores and 2 GPUs as accelerator cards, a simulation with $256^3$ mesh cells could be run by assigning slabs with a thickness of 98 cells to each of the GPUs, while letting the remaining 30 compute cores each work on slabs with a thickness of 2 cells each. Of course, this particular  mixed execution example would only make sense if each of the GPUs would be around $\sim 50$ times faster than a single CPU core. In practice, the speed difference is typically considerably larger, so most of the work should typically be assigned to GPUs if those are available.

We also note that for the moment our code supports only meshes with uniform and fixed resolution. However, a more general domain decomposition than just a slab-based decomposition is planned for the future and in principle straightforward. This can, in particular, remove the obvious scaling limitation of our current approach, where the number of cells per dimension sets the maximum number of GPUs or CPU cores that could be employed.

\subsection{GPU computing implementation}

The above parallelization strategy makes it clear that our code is neither a plain CPU code nor a pure GPU code. Rather, it implements its core compute functionality where needed twice, in a CPU-only version and in a GPU-only version. Both versions can be interchangeably used for any given slab taken from the global computational mesh, and they produce the same results. While this approach evidently requires some extra coding, we have found that this is actually quite helpful for code validation, as well as for quantifying the relative performance of CPU and GPU versions. Further, the extra coding effort can be greatly alleviated by using wherever possible functions that can be compiled and executed both by GPUs and CPUs based on a single implementation.

For the GPU code, we have used the CUDA programming model available for Nvidia GPU devices. All our code is written in low level C++, and we presently do not make use of programming models such as OpenACC, special GPU-accelerated libraries, or new C++ language features that allow GPU-based execution of standard libraries via execution policies. Our programming model is thus best described as MPI-parallel C++, accelerated with CUDA\footnote{We presently use the CUDA toolkit version 11.4, the GNU g++ 11 compiler and the C++17 standard. For message passing, we prefer the OpenMPI-4 library, for Fourier transforms we use FFTW-3 and for random number generation we rely on GSL 2.4.} when GPUs are available. If no GPUs are available, the code can still be compiled into a CPU-only version.

For storing static data such as coefficients of Legendre polynomials or Gaussian quadrature weights, we try to make use of the special constant memory on GPUs, which offers particularly high performance, also in comparison to the ordinary general memory. Likewise, for computing parallel reductions across individual cells, we make use of the special shared memory. However, the size of the corresponding memory spaces is quite limited, and varies between different GPU hardware models. This can necessitate adjustments of the used algorithms at compile time, depending on code settings such as the expansion order and  on which execution platform is used. We address this by defining appropriate compile-time switches, such that these adjustments are largely automatic.

We note that the data of slabs that are computed with GPUs need to fit completely on GPUs as we refrain from transmitting the data from the front end host computer to the GPU on every timestep. Instead, the data remains on the GPU for maximum performance, and only when a simulation is finished or a temporary result should be output to disk it is copied back from the GPU to the front end host. Wherever such transfers are needed, we use pinned memory on the front end to achieve maximum bandwidth between the host and GPUs. GPUs can access such pinned memory directly, without going through the host CPU first. The problem sizes we are able to efficiently tackle with GPUs are therefore limited by the total combined GPU memory available to a run. Modern GPUs  typically have some 10~GBs of main memory, but the detailed amount can vary greatly depending on the model, and is course a matter of price as well. The communication between adjacent slabs is organized such that communication and computation can in principle overlap. This is done such that first the surface states are computed and a corresponding MPI exchange with the neighbouring slabs is initiated. While this proceeds, the volume integrals for slabs are carried out by the GPU, and only once this is completed, the work continues with the received surface data.

Because slabs that are computed on GPUs need to be executed in a massively thread-parallel fashion with shared-memory algorithms, some changes in the execution logic compared to the effectively serial CPU code are required. For example, to avoid race-conditions in our GPU code without needing to introduce explicit locks, we process the mesh in a red-black checkerboard fashion. Finally, we note that we also implemented a scheme that makes our results binary identical when the number of mesh slabs is changed. This ultimately relates to the question about how the wrap-around between the leftmost and rightmost planes of the mesh in our periodic domain is implemented. Here the order in which fluxes from the left and right neighboring cells is added to cells needs to be unique and independent of the location and number of slabs in the box in order to avoid that different floating point rounding errors can be introduced when the number of slabs is changed.

\begin{table}
\begin{center}
\begin{tabular}{lrc}
\hline
    $N_c$  &  $p$ &  min. memory need \\
    \hline
 128  &  1 &  512 MB\\ 
 128  &  2 &  1440 MB\\
 128  &  3 &   3520 MB\\ 
 128  &  5 &   9856 MB\\
 128  &  9 &  37.81 GB \\ 
 2048  &  1 &  2048 GB\\ 
 2048  &  2 &  5760 GB\\
 2048  &  3 &   13.75 TB\\ 
 2048  &  5 &   38.5 TB \\
 2048  &  9 &  151.3 TB \\ 
     \hline
\end{tabular}
\end{center}
    \caption{Minimum memory need for our DG code when a 3D simulation is assumed with $(N_\textrm{c})^3$ cells and expansion order $p$, including allowing for an  artificial viscosity field. Here double precision with 8 bytes per floating point number has been assumed. }
    \label{tab:mem}
\end{table}

\begin{figure*}
	\resizebox{18cm}{!}{\includegraphics{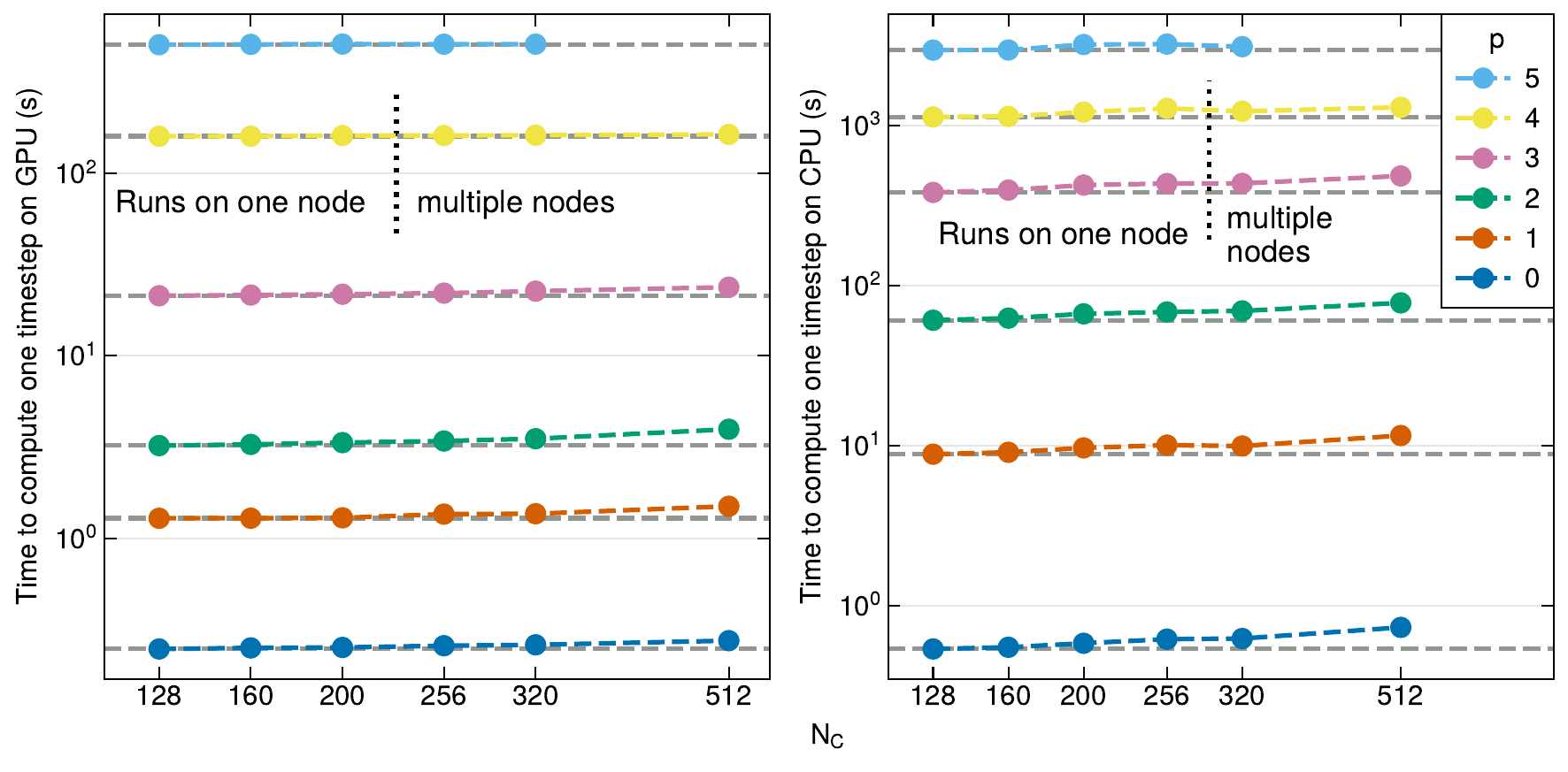}}
    \caption{Weak scaling of {\small TENETGPU} for a 3D test problem. The $y$-axis shows the time taken to compute one timestep averaged over a small number of timesteps. The left panel shows results for GPU execution when the problem size $N_c^3$, measured in terms of the number of cells $N_c$ per dimension, increases in several steps by close to a factor of two from $N_c=128$ to $N_c=512$ cells, and when between 1 to 64 GPUs are applied to the problem. In contrast, the right hand panel gives results when the problems are executed on CPUs instead, using from 4 to 256 cores, again keeping in each case the load per computational element constant. We carry out the measurements for different expansion order, from $p=0$ to $p=5$. Ideal weak scaling corresponds to horizontal lines (dashed). The dotted vertical line marks the transition between using CPU cores or GPUs associated with a single compute node of our cluster, and the use of multiple nodes in which MPI data exchange via the Intel Omni-Path takes place. The missing measurement at $p=5$ is due to the large memory required to store communication buffers, which make the $N_\textrm{c}=512$ problem not fit onto 64 GPUs. The missing data points at $N_c=400$ are due to 400 not being divisible by 32, as this would lead to uneven distribution of work across the GPUs we did not consider these runs.}
    \label{fig:weak_scaling}
\end{figure*}

\begin{figure*}
    \resizebox{18cm}{!}{\includegraphics{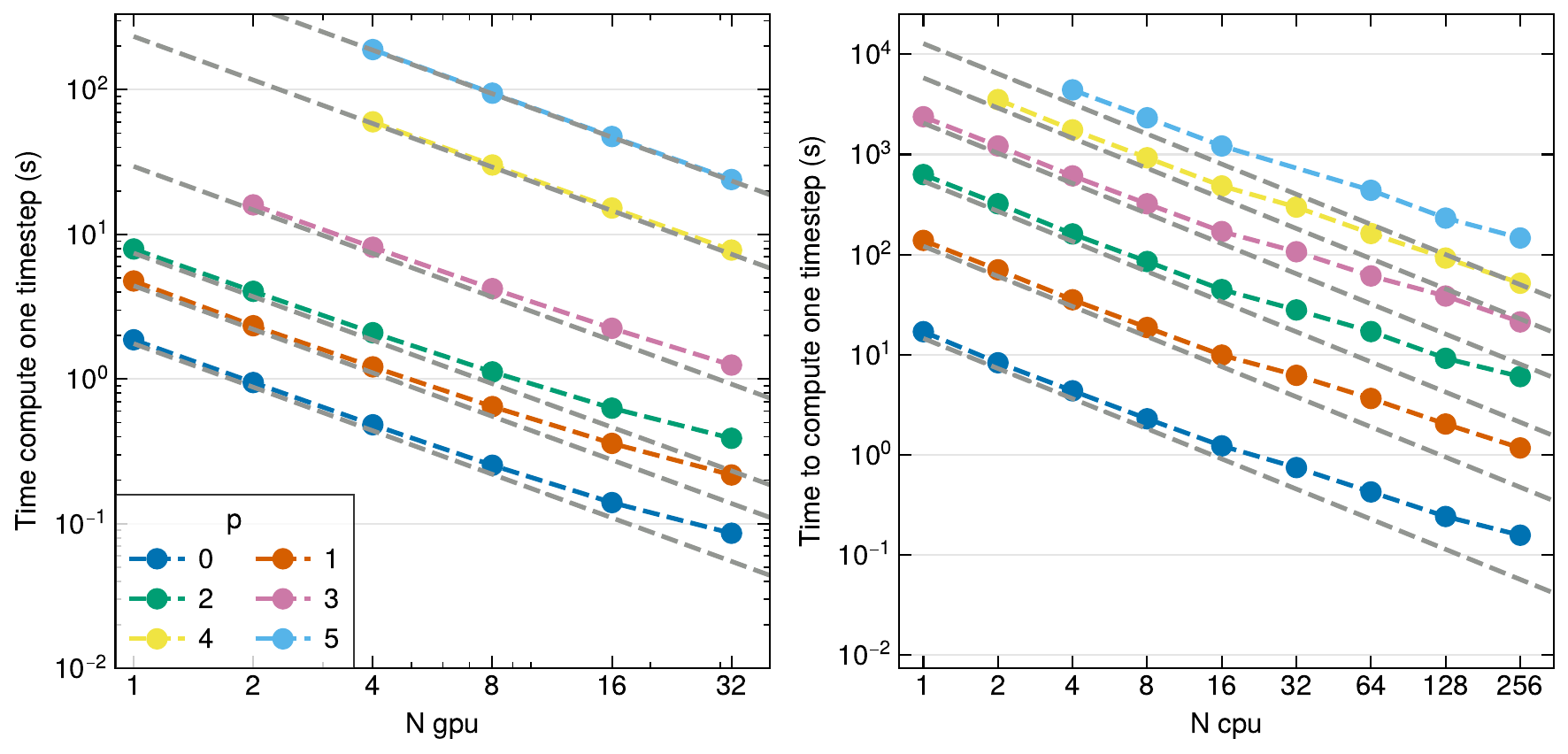}}
    \caption{Strong scaling of {\small TENETGPU} for a 3D test problem of size $256^3$ cells. The $y$-axis shows the average time taken to carry out one timestep. The left panel shows timing results when between 1 and 16 Nvidia A100 GPUs are used, while the right panel gives results when between 1 to 256 ordinary Intel Xeon-6138 cores are used. Ideal strong scalability corresponds to the dashed lines indicated in the panels. Missing data points at high orders and low number of compute devices are due to the fact that such large problems do not fit on a single GPU / node.}
    \label{fig:strong_scaling}
\end{figure*}

\subsection{Memory usage}

Before closing this section, it is perhaps worthwhile to discuss the memory need of our DG simulations, as this is ultimately determining the maximum size of simulations that can be done for a given number of GPUs. To represent a scalar field such as the density $\rho$ at order $p$, we need for every cell a certain number of basis function weights $N^{d{\rm D}}(p)$, were $d$ is the number of spatial dimensions, see equations (\ref{eqn:N3D}) and (\ref{eqn:N2D}). When multiplied with the number of cells, we obtain the number of degrees of freedom, which is identical to the number of floating point variables needed to stored the full density field. If we write the total number of cells as $(N_c)^d$, then the total number of variables that need to be stored for the DG weight vector is
\begin{equation}
N_w = (2+d)  (N_c)^d N^{d{\rm D}}(p).
\end{equation}
Here we assumed that we simulate the plain Euler equations without viscosity, where we need $(2+d)$ conserved fields to describe the flow. If we account for our artificial viscosity field, which will always be required for problems involving shocks, this number goes up by one further unit, yielding
\begin{equation}
N_w = (3+d)  (N_c)^d N^{d{\rm D}}(p).
\end{equation}
A passive tracer field, if activated, would add a further unit in the prefactor. In 2D and 3D, a conservative upper bound for $N^{d{\rm D}}(p)$ is $p^d$, but this is not particularly tight. Already for $p=2$, 
$N^{3{\rm D}}$ is lower than $p^3$ by a factor of 2, for $p=4$ this grows to a factor 3.2, and for $p=10$ the difference is  more than a factor 4.5.

Another significant source of memory need lies in our timestepping algorithm. At present we use stability preserving Runge-Kutta schemes that require a temporary storage of the time derivatives of the weights, evaluated at several different points in time, depending on the order of the Runge-Kutta scheme, which we adjust according to the chosen $p$. The required temporary storage space $N_{\dot{w}}$ is thus a multiple of $N_w$, with a prefactor that depends on the chosen order $p$, i.e.
\begin{equation}
N_{\dot{w}} = f_t(p) N_w.
\end{equation}
Here $f_t(p)$ depends on the number of stages in the Runge-Kutta scheme. Presently, we use a setup where $f_t(p) = p$ for $p \le 3$, and $f_t(p) = 5$ otherwise. The minimum amount of total storage (in terms of needed floating point numbers) required by the code is thus
\begin{equation}
N_w = [3+d + f_t(p)]  (N_c)^d N^{d{\rm D}}(p).
\end{equation}

During execution of our code using multiple GPUs or CPU cores, some temporary buffer space is furthermore required to hold, in particular, send and receive buffers for fluid states or fluxes along slab surfaces orthogonal to the $x$-direction. These tend to be subdominant, however, compared to the memory requirements to store the weights and their time derivatives themselves. The latter thus represent the quantities that need to be primarily examined to decide about the feasibility of a simulation in terms of its memory needs. When we use the oscillatory sensor for controlling artificial viscosity, some further temporary storage is needed as well, but since this is again small compared to $N_w$ since only two scalar quantities per cell are needed, this conclusion is not changed. Note that our DG approach does not need to store gradient fields for any of the fields, which is different from many finite volume methods such as, for example, {\small AREPO}. Also, use of the Navier-Stokes solver instead of simulating just the Euler equations does not increase the primary memory needs in any significant way.

In Table~\ref{tab:mem}, we give a few examples of the memory need for a small set of  simulation sizes and simulation orders, which illustrates the memory needs of the code, and which can be easily scaled to other problem sizes of interest. A single Nvidia A100 GPU with 40 GB of RAM could thus still run a $N_c=128$ problem at order $p=9$, or a  $512^3$ problem at quadratic order $p=1$. For carrying out a $2048^3$ simulation at $p=1$, a cluster offering at least 52 such devices would already be necessary.

\section{Code performance}
\label{SecPerformance}

In order to fully utilise large parallel supercomputers, a code has to be able to run efficiently not only on a single core on one CPU, but also on hundreds to thousands of cores on many CPUs. The degree to which this can accelerate the total runtime of a computation is encapsulated by the concept of parallel scalability. Similarly, for a GPU-accelerated code it is of interest to what extent the use of a GPU can speed up a computation compared to using an ordinary CPU. If more than a single GPU is used, one is furthermore interested in whether a code can efficiently make simultaneous use of several, perhaps hundreds of GPUs. In this section we examine these aspects and present results of weak- and strong scaling tests of our new code.

\subsection{Weak scaling}

Weak scaling performance describes a situation where a set of simulations of increasing size is run and compared, but where the load per computational unit, be it a CPU core or a GPU in our case, is kept constant. The time to perform a single timestep should remain constant in this case, increasing only due to communication-related overhead, through work-load imbalances, or through other types of parallelization losses, for example if a code contains residual serial work that scales with the problem size. 

Weak scaling results of our code are shown on Fig.~\ref{fig:weak_scaling}. We run a 3D box with constant density and pressure using the Navier-Stokes equations, the positivity limiter and artificial viscosity. This setup is computationally very close to problems we are running in production. We consider problem sizes of $128^3$, $160^3$, $200^3$, $256^3$, $320^3$, and $512^3$ cells, forming a sequence that approximately doubles in size, with a factor of 64 enlargement from the smallest to the largest runs. To compensate for the fact that the problem size does not exactly double every time we increase the number of cells, we apply a correction factor to the timing results at each resolution\footnote{The current version of the GPU part of the code can only run if $N_\textrm{c}$ and the number of slabs in the $x$-direction per rank are even. This and the fact that $N_\textrm{c}$ has to be an integer in any case prevents ideal doubling of problem size. The correction factors we apply are: 128$^3$: 1.0, 160$^3$: 0.977, 200$^3$: 0.954, 256$^3$: 1.0, 320$^3$: 0.977, and 512$^3$: 1.0.}. Correspondingly, we execute these problems with one \mbox{Nvidia A100 GPU} for the smallest mesh size, and 64 GPUs for the largest mesh size, keeping the load per GPU roughly constant. The results are shown in the left panel of Fig.~\ref{fig:weak_scaling}. For comparison, we also measure the execution speed if instead every GPU is replaced by four CPU cores of Intel Xeon-6138 processors. The corresponding results are shown in the right panel of  Fig.~\ref{fig:weak_scaling}. Finally, we repeat these measurements for different DG expansion orders $p=0-5$.

The results in the figure show generally good weak scalability, but also highlight some performance losses for large problem sizes. These arise in part because our domain is split into slabs and not cubes. Larger problems lead to ever thinner slabs with a larger surface-to-volume ratio and thus more communication between different slabs. We also see the influence of enhanced communication on weak scalability when data needs to be transferred across node boundaries. At higher orders the weak scaling is generally better, as the compute-to-communicate time ratio shifts strongly to the compute side.

\subsection{Strong scaling}

Strong scaling is a test where one runs a problem of given size on an ever increasing number of compute units. Contrary to weak scaling, the load per compute unit decreases in this test, and  the time to perform a single timestep should decrease in inverse proportion to the increasing computational power applied to solve the problem. 

We show a strong scaling result in Fig.~\ref{fig:strong_scaling}, again carried out for a 3D box with constant density and pressure using the Navier-Stokes equations, the positivity limiter and artificial viscosity. For definiteness, we use a simulation with $256^3$ cells, and consider orders $p=0$ to $p=5$. The left panel of Figure~\ref{fig:strong_scaling} shows the average execution time for a single step when 1, 2, 4, 8, or 16 Nvidia A100 GPUs are used. In contrast, the right panel of Figure~\ref{fig:strong_scaling} shows the average execution time when CPU cores on a cluster with 2 Intel Xeon-6138 CPUs are used, with 40 cores per node. We show results from 1 core to 256 cores. Especially in the latter case, one sees clear limits of strong scalability, as communication costs become quite large if the problem is decomposed into slabs that are just a single cell wide. This stresses that there is always a limit for strong scalability, something that is known as Ahmdahl's law. By enlarging the problem size, this limit can however usually be pushed to larger parallel partition sizes. Another major contributor to the degradation that happens when going from 16 to 32 cores is the saturation of available memory bandwidth of a single 20-core socket. We verified this using the STREAM benchmark\footnote{\url{https://github.com/intel/memory-bandwidth-benchmarks}}.

\subsection{CPU vs GPU benchmark}

Another interesting question is how the absolute speed of GPU execution of our code compares to running it only on ordinary CPU cores. To estimate this speedup we take the average execution times to compute a timestep from our weak scaling results for both the GPU and CPU runs and consider their ratio. We do this for the three considered DG orders $p=2$ to $p=4$, and for the varying problem sizes and number of compute units used. Since we had used 4 CPU cores to pair up with 1 GPU, we rescale the results in two different ways, to either compare the execution performance of four Nvidia A100 GPUs with 40 Intel Xeon-6138 cores -- which is how one of our compute nodes is equipped -- or to the performance of a single GPU compared to one CPU core (which thus gives 10 times higher values).

\begin{figure}
    \includegraphics[width=88mm]{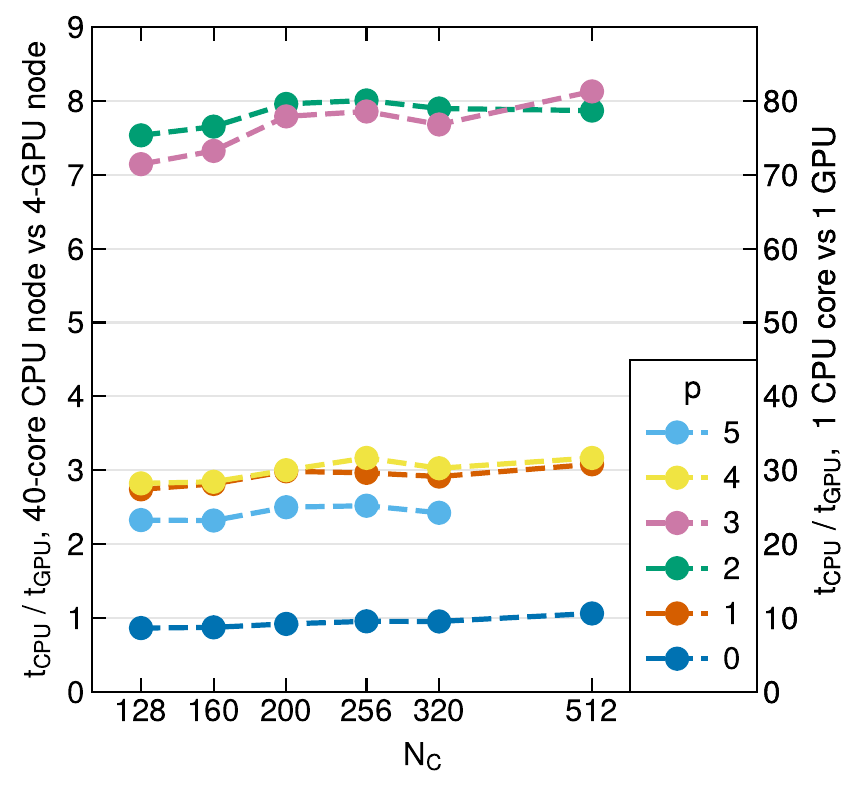}
    \caption{Ratio of time taken to calculate one timestep of test simulations with the Navier-Stokes solver on GPUs or CPUs, based on our weak scaling test runs. The left vertical scale shows results when we normalize them to the speed ratio for using 4 Nvidia A100 GPUs versus 40 Intel Xeon 6138 CPU cores, while the right scale normalizes the speed results to a comparison of 1 GPU vs 1 CPU core.}
    \label{fig:gpu_speedup}
\end{figure}

The corresponding results are illustrated in Fig.~\ref{fig:gpu_speedup}. The speedup of GPU execution at the node-level is modest for order $p=2$, as there are not enough floating point operations to fill up the GPUs. At $p=2,\ 3$ we reach the highest node-level speedup observed among this set of runs, it peaks at just over 8x the CPU speed for large problems. This runs show better performance because there are a lot of floating points operations to perform at the same time, and all intermediate results still fit into the GPU's limited shared memory. Such shared memory is ``on chip'' and therefore about $\sim$100x faster than global memory. Once the intermediate results become too large to fit into shared memory, the code determines the maximum number of quadrature points it can process at once and proceeds forward in batches of $n$ quadrature points. At this point, a single GPU is about 80 times as fast as a CPU core, but when comparing a fully equipped GPU node to a fully equipped CPU node, more realistic numbers are in the ballpark of $\sim 8$. Note that this speedup metric is based on the specific hardware configuration of the cluster the authors had access to throughout this project. While the configuration of four Nvidia A100 GPUs paired with about 40 Intel Xeon cores quite typically reflects the general HPC situation in 2021 and 2022, the corresponding hardware characteristics are not universal and can be expected to evolve substantially in future generations of CPU-GPU systems. In any case, the performances we find are not far away from the ratio of the nominal peak performances of the involved compute devices for double precision arithmetic (which we have used here throughout), but this comparison also suggests that there is still some modest room for improvement in the performance of our GPU implementation. 

\section{Summary and Conclusions}
\label{SecSummary}

In this study, we have described a novel hydrodynamical simulation code which is based on the mathematical  Discontinuous Galerkin  approach. The fluid state is expanded in this method into a set of spatially varying basis functions with time-variable weights, yielding a separation of the temporal and spatial  dependencies. The time evolution of the weights is obtained in a weak formulation of the underlying partial differential equations of fluid dynamics. 

Our work builds up on the earlier development of a DG code by \citet{Schaal2015} and \citet{guillet_high_order_2019}, but extends it into several crucial directions. First of all, we have developed a novel GPU implementation from scratch, thereby demonstrating the substantial potential of these acceleration devices for achieving higher computational performance in astrophysical applications. This potential has already been identified in a few first finite-volume hydrodynamical GPU codes in astrophysics, but ours is the first one that can carry out DG calculations of the full Navier-Stokes equations at very high order of $p= 10$ and beyond.

Secondly, we have introduced a novel approach to shock-capturing at high order, solving the long-standing problem that standard slope-limiting techniques do not work well at high order and tend to discard in troubled cells much of the advantage that is supposed to be delivered by a high order approach. The latter can only be rescued if the DG method is able to capture physical discontinuities in a sub-cell fashion. By means of our new source routines for a time-dependent artificial viscosity field, we have demonstrated very good shock-capturing ability of our code, with a shock broadening that closely tracks the effective spatial resolution $h/p$ that we expect from the method based on its number of degrees of freedom per dimension. While this does not necessarily give high-order approaches an advantage for representing a shock compared with a lower order method with the same number of degrees of freedom, at least it also is not worse -- using a high-order approach will however in any case still be beneficial for all smooth parts of a flow. If it performs at the same time as well as a lower order method in places where there is a shock, this can be a significant advantage. For contact discontinuities, similar considerations apply, but here high-order methods have the additional advantage of exhibiting greatly reduced numerical diffusivity. Contact discontinuities that move over substantial timespans therefore also benefit from the use of higher order.

Third, we have stressed that the use of physical viscosity is often a key to make problems well posed and amenable to direct numerical solutions. Here we have introduced a  novel method to define the viscous surface fluxes at cell interfaces. This is based on arriving at unambiguous derivatives at interfaces by projecting the two piece-wise solutions in the adjacent cells onto a continuous basis function expansion covering both cells. The derivatives can then be computed in terms of analytic derivatives of the basis functions. We have shown that this technique is robust, consumes much less memory and computational effort than the uplifting technique, and most importantly, it converges at the expected rapid convergence rate when high order is used.

In fact, in several of our test problems, we could show that our DG code shows for smooth problems exponential convergence as a function of expansion order $p$, while for fixed order, the $L_1$ error norm declines as a power-law of the spatial resolution, $L_1\propto h^{p}$. These favourable properties suggest that it is often worthwhile to invest additional degrees of freedom into the use of higher expansion order rather than employing more cells. However, since every DG cell effectively represents a small spectral problem in which the required solution evaluations and volume integrations are carried out in real space, the computational cost to advance a single cell also increases rapidly with order $p$. In practice, this can make the optimal order quite problem dependent.

With our present implementation we could obtain excellent agreement with the reference Kelvin-Helmholtz solution computed by \citet{Lecoanet2016} with the spectral code {\small DEDALUS}. Remarkably, we achieved this already with 64 cells and order $p=4$, for which our results are equally as accurate as those obtained with the finite volume code {\small ATHENA} at second order using 2048 cells. This again shows the potential of the DG approach. Given that in this work we could overcome one of its greatest weaknesses in an accurate, simple, and robust way -- namely the treatment of shocks at high order -- we are confident that the DG method could soon turn into a method of choice in astrophysical applications, rivaling the traditional finite volume techniques. Our next planned steps to make this a reality are to add additional physics such as radiative cooling and self-gravity to our code, and to provide functionality for local refinement and derefinement ($h$-adaptivity), as well as to allow for varying the expansion order used in a single cell ($p$-adaptivity). The high performance we could realize with our GPU implementation, which outperforms modern multi-core CPUs by a significant factor, furthermore strengthens the case to push into this direction, which seems also a necessity to eventually be able to harness the power of the most powerful supercomputers at the exascale level for unsolved problems in astrophysical research.

\section*{Acknowledgements}

The authors would like to thank the anonymous referee whose comments significantly improved the quality of this work. We also thank Philipp Grete for discussions about the bottleneck effect and to Damien Begue for insights about recovery based advanced DG methods. The authors also acknowledge helpful discussions with numerous students at MPA, this pandemic-era work could not be finished without your support.\vspace*{-0.5cm}


\section*{Data Availability}

Data of specific test simulations can be obtained upon reasonable request from the corresponding author.\vspace*{-0.5cm}



\bibliographystyle{mnras}
\bibliography{references}

\bsp	
\label{lastpage}
\end{document}